\documentclass[twocolumn,letterpaper]{article}

\usepackage{bm}
\usepackage{subfigure}

\ifx\pdfoutput\undefined
\usepackage{graphicx}
\else
\usepackage[pdftex]{graphicx}
\usepackage{epstopdf}
\fi

\usepackage{amssymb,amsmath}
\usepackage{authblk}

\title{Dynamics of Oscillators Coupled by a Medium\\with Adaptive Impact}

\author{Roozbeh Daneshvar\\
  Department of Aerospace Engineering\\
  Texas A\&M University\\
  \texttt{roozbeh@tamu.edu}}

\date{}

\begin{document}

\maketitle


\begin{abstract}
In this article we study the dynamics of coupled oscillators. We use mechanical metronomes that are placed over a rigid base. The base moves by a motor in a one-dimensional direction and the movements of the base follow some functions of the phases of the metronomes (in other words, it is controlled to move according to a provided function). Because of the motor and the feedback, the phases of the metronomes affect the movements of the base while on the other hand, when the base moves, it affects the phases of the metronomes in return.

For a simple function for the base movement (such as $y = \gamma_{x} [r \theta_1 + (1 - r) \theta_2]$ in which $y$ is the velocity of the base, $\gamma_{x}$ is a multiplier, $r$ is a proportion and $\theta_1$ and $\theta_2$ are phases of the metronomes), we show the effects on the dynamics of the oscillators. Then we study how this function changes in time when its parameters adapt by a feedback. By numerical simulations and experimental tests, we show that the dynamic of the set of oscillators and the base tends to evolve towards a certain region. This region is close to a transition in dynamics of the oscillators; where more frequencies start to appear in the frequency spectra of the phases of the metronomes.
\end{abstract}


\section{Introduction}


%
Dissipative systems with a nonlinear time-delayed feedback or memory can produce chaotic dynamics \cite{MG-OCPCS-1977,F-CAIDDS-1982}. The effect of the delay on the dimension of these chaotic attractors is shown in \cite{DGLPRT-SDCDDS-1987}. Delay systems generically have families of periodic solutions, which are reappearing for infinitely many delay times. As delay increases, the solution families overlap leading to increasing coexistence of multiple stable as well as unstable solutions \cite{YP-DP-2009}. Anticipating chaotic synchronization is discussed in \cite{V-ACS-2000}.
%


Packard \cite{P-ATEC-1988} showed adaptation to the edge of chaos in cellular automata rules with genetic algorithms. Some of his results were later disputed in \cite{MHC-REC-1993}. Co-evolution to the edge of chaos is discussed in \cite{KJ-CECCFLPSCA-1991}. Edge of chaos has been found to be the optimal setting for control of a system \cite{PH-TACALMD-1994}. A self-adjusting system is a system in which the control of a parameter value depends on previous states of the system \cite{HW-SASAC-2009}. The authors in \cite{MKWH-AECSALM-2000} describe adaptation to the edge of chaos in logistic map. They believe that adaptation to the edge of chaos is a generic property of the systems with a low-pass filtered feedback.They believe that this property is independent of the form of the feedback and the system under study. The findings have also been confirmed experimentally with Chua's circuit \cite{CKM-DSF-1986}. In \cite{BH-CQAEE-2006} conserved quantities are used for investigating adaptation to the edge of chaos. The phrase \emph{Edge of Chaos} was originally proposed by Chris Langton in 1990 in the area of cellular automata \cite{L-CECPTEC-1990} although some others mentioned the same concept around the same time \cite{CY-COC-1990}. Guiding an adaptive system through chaos is also considered in \cite{HP-GASTC-2007}. Adaptation to the edge of chaos is studied in \cite{AECODCM-AA-2006}. In that research, one-dimensional chaotic maps are used as dynamical systems and feedback of values obtained from observations of the system variable are provided for the system. The parameter changes more slowly than the variable. The authors believe that separation of time scales is necessary for the method to be self-contained without external control. The parameter governs the dynamics of the variable every certain number of steps of time series. On the other hand, the variables of the system determine the dynamics of the parameter. Extinction in a simple ecological model is shown in \cite{SP-UDESEM-1996}. It is stated that chaotic dynamics does not necessarily lead to population extinction. The topic is also discussed in \cite{K-OO-1993,BN-RTCECRNN-2004,SMS-ECCMMVHL-2004,MCH-DCECRE-1994}.
%
%
\subsection*{Purpose of this Research}
We are studying coupled oscillators. We study the case in which we interfere with the coupling and for this case, we use the concept of forced oscillators and use mechanical metronomes placed on a rigid base. We study the effects of movements of the base on the dynamics of the oscillators. This makes the whole set a dynamical system which contains two sub-systems (the oscillators) that has many potentials to be explored. We will consider the case where the system can adjust itself; In other words, we study how the system evolves in time when some parameters of the system are adjusted by states of the system itself.
%

\section{Forced Oscillators}


In this section we investigate the effect of forcing the oscillators on the dynamics. To read more about details of the experiments, please see Appendix \ref{sec-Experiments} and to read more about synchronization of oscillators when the base is freely moving, please see Appendix \ref{sec-synchronization}.


\subsection{Forced Oscillators}

When the metronome base moves, we have a forced oscillator for which the governing equation is as below \cite{P-SM-2002}
\begin{equation}
\begin{split}
\frac{d^2 \theta}{d t^2} + \frac{m r_{c.m.} g}{I} \sin{(\theta)} + \epsilon \left[\left(\frac{\theta}{\theta_0}\right)^2 - 1\right] \frac{d \theta}{d t}
\\
+ \underbrace{\left(\frac{m r_{c.m.}}{I}\right)}_{\frac{k}{g}} \cos{(\theta)} \frac{d^2 x}{d t^2} = 0
\end{split}
\end{equation}
in which $x$ is the horizontal position of the base. We write the governing equation as below
\begin{subequations}
\begin{align}
\frac{d \theta}{d t} &= \omega \\
\frac{d \omega}{d t} &= - k \sin{(\theta)} - \epsilon \left[\left(\frac{\theta}{\theta_0}\right)^2 - 1\right] \omega - \frac{k}{g} \cos{(\theta)} \frac{d^2 x}{d t^2}
\end{align}
\end{subequations}
When we form a feedback loop for the movements of the base, we have the following set of equations
\begin{subequations}
\begin{align}
\frac{d \theta}{d t} &= \omega \\
\frac{d \omega}{d t} &= - k \sin{(\theta)} - \epsilon \left[\left(\frac{\theta}{\theta_0}\right)^2 - 1\right] \omega - \frac{k}{g} \cos{(\theta)} z \\
\frac{d z}{d t} &= h(\theta , \gamma)
\end{align}
\end{subequations}
in which
\begin{equation}
z = \frac{d^2 x}{d t^2}
\end{equation}
and $h(\theta , \omega)$ is a function specifying the feedback. This function depends on how the base responds to the states of the system such as the detected angles of the metronomes. For instance, $h(\theta , \omega) = \theta$ means that the cart sets its acceleration to the same as the phase of the oscillator.


\subsection{Two Oscillators on a Controlled Base}

In this case we consider two metronomes placed on a moving rigid base. The metronomes oscillate freely while the movements of the base impacts them. The equations of motion for this case are as follows

\begin{subequations}\label{eq:MainEqTwoMetsOneActiveBase}
\begin{align}
\dot{\theta}_1 &= \omega_1 \label{eq:MainEqTwoMetsOneActiveBaseFirst}\\
\dot{\omega}_1 &= - k_1 \sin{(\theta_1)} - \epsilon \left[\left(\frac{\theta_1}{\theta_0}\right)^2 - 1\right] \omega_1 - \frac{k_1}{g} \cos{(\theta_1)} z \label{eq:MainEqTwoMetsOneActiveBaseSecond}\\
\dot{\theta}_2 &= \omega_2 \label{eq:MainEqTwoMetsOneActiveBaseThird}\\
\dot{\omega}_2 &= - k_2 \sin{(\theta_2)} - \epsilon \left[\left(\frac{\theta_2}{\theta_0}\right)^2 - 1\right] \omega_2 - \frac{k_2}{g} \cos{(\theta_2)} z \label{eq:MainEqTwoMetsOneActiveBaseFourth}\\
\dot{z} &= h(\theta_1, \omega_1, \theta_2, \omega_2) \label{eq:MainEqTwoMetsOneActiveBaseFifth}
\end{align}
\end{subequations}
in which $\theta_1$ and $\theta_2$ are the angles the metronomes make with the vertical and $\omega_1$ and $\omega_2$ are their derivatives (i.e. rotational velocities). $x$ is the horizontal position of the base, $y = \dot{x}$ is the linear velocity of the base and $z = \dot{y}$ is the acceleration of the base (we have only included $z$ because it is the only parameter affecting the oscillators). $I$ is the moment of inertia of the pendulum, m is the mass of the pendulum, $r_{c.m.}$ is the distance of the pendulum’s center of mass from the pivot point and $g$ is the acceleration of gravity.
%

\subsection{$y = \gamma_x (r \theta_1 + (1 - r) \theta_2)$}


As the servo motor in the experiments receives desired velocity as the input (and not the acceleration or first derivative of acceleration), we design the function for change of $z$ based on a function for $y$. For this case we consider that
\begin{equation}
y = \gamma_x (r \theta_1 + (1 - r) \theta_2)
\end{equation}
so
\begin{equation}
z = \dot{y} = \gamma_y (r \dot{\theta}_1 + (1 - r) \dot{\theta}_2)
\end{equation}
and
\begin{equation}
\begin{split}
\dot{z} &= \gamma_z (r \ddot{\theta}_1 + (1 - r) \ddot{\theta}_2)
\\
        &= \gamma_z (r \dot{\omega}_1 + (1 - r) \dot{\omega}_2)
\end{split}
\end{equation}
A sample of frequencies for a range of ratios is shown in Fig. \ref{fig:FS-for-Theta}.
\def \picwidth {0.47}
\begin{figure*}[htb]
\centering
\subfigure[Frequency spectrum of $\theta_1$ for $k_1 = 6.3$ and $k_2 = 6.3$]{
\includegraphics[width=\picwidth\textwidth]
{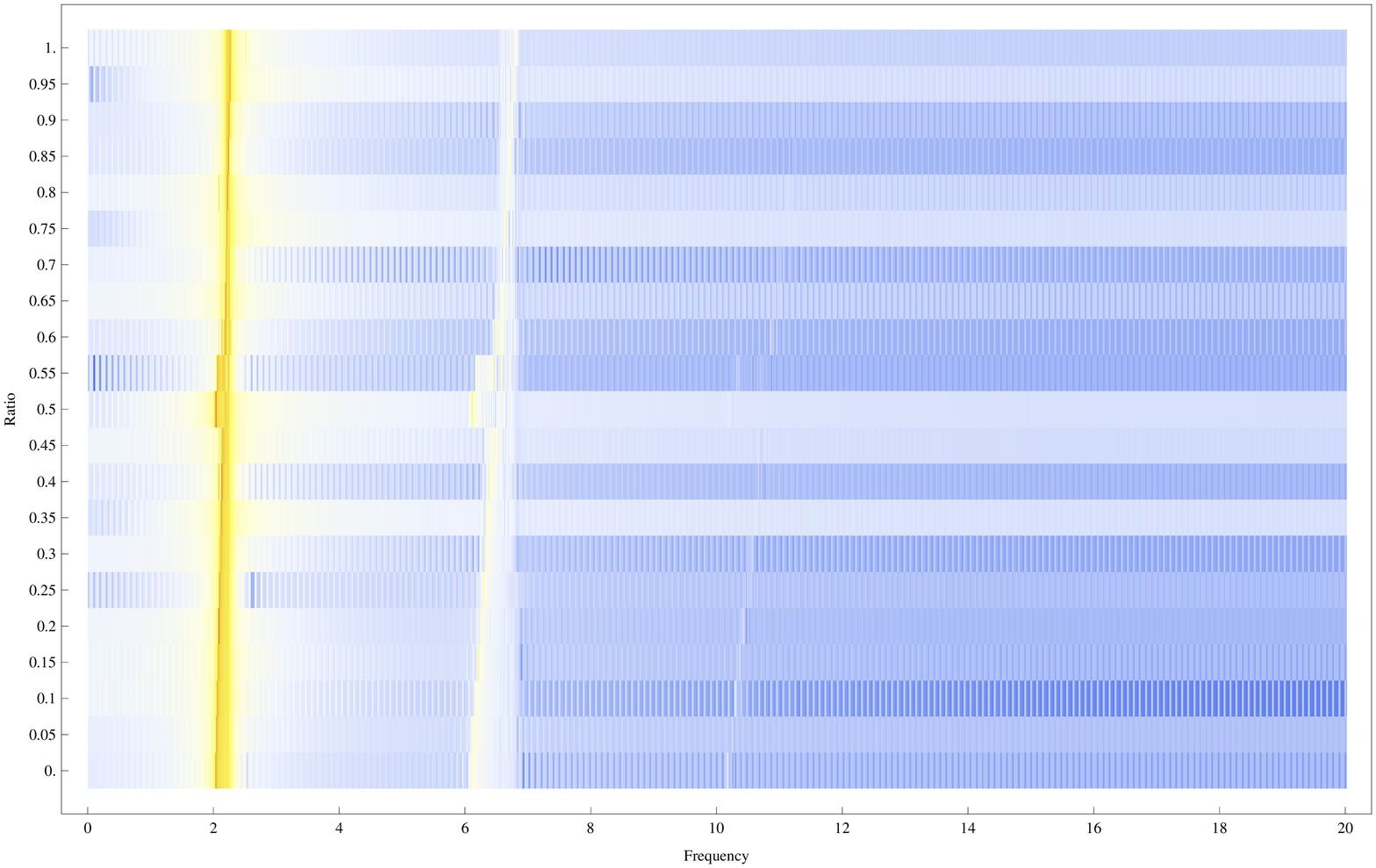}
}
\label{fig:FS-Theta1-k1-6-3-k2-6-3}
\subfigure[Frequency spectrum of $\theta_2$ for $k_1 = 6.3$, $k_2 = 6.3$]{
\includegraphics[width=\picwidth\textwidth]
{FS-Theta1-k1-6-3-k2-6-3.eps}
}
\label{fig:FS-Theta2-k1-6-3-k2-6-3}
\subfigure[Frequency spectrum of $\theta_1$ for $k_1 = 6.3$, $k_2 = 18.7$]{
\includegraphics[width=\picwidth\textwidth]
{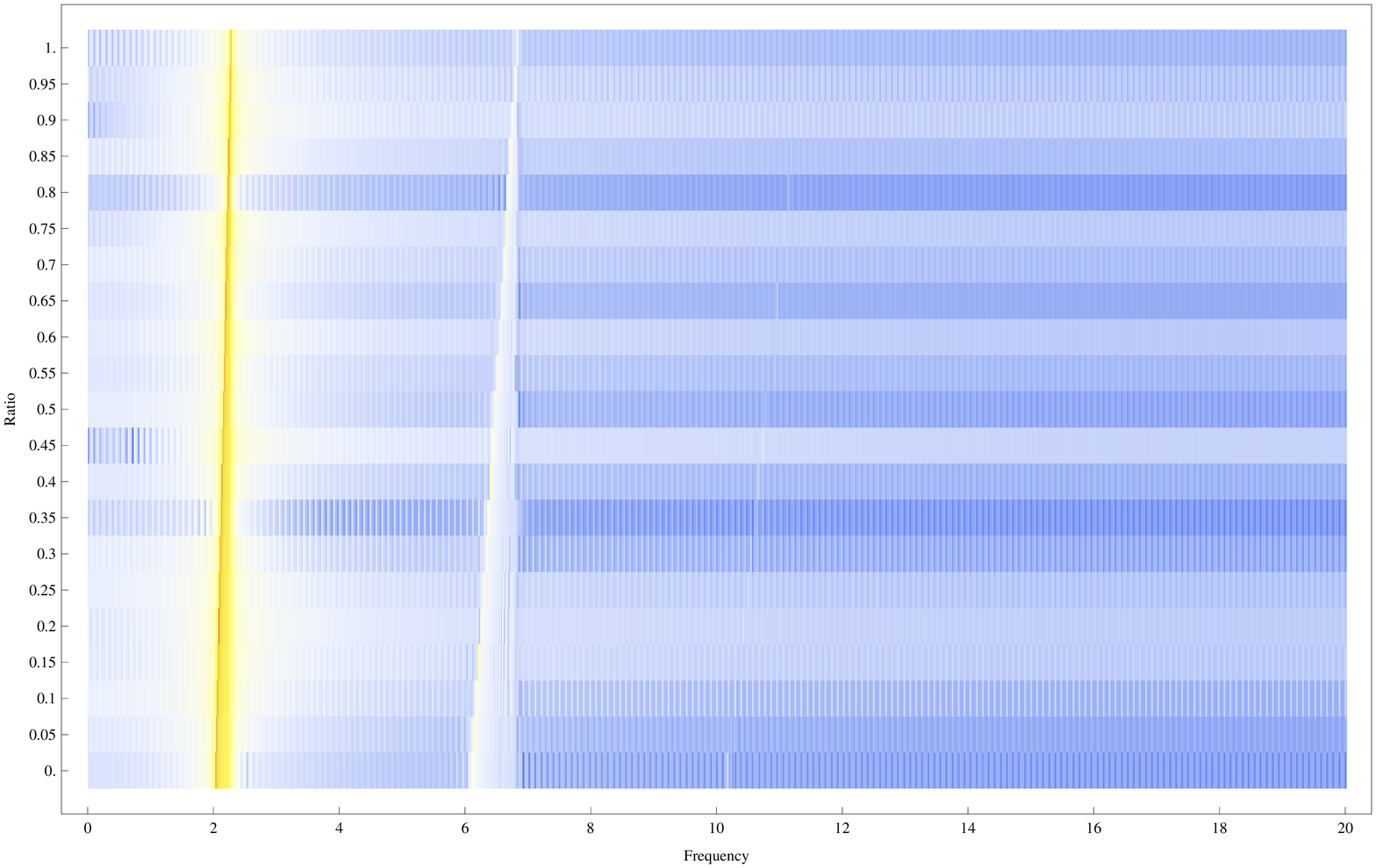}
}
\label{fig:FS-Theta1-k1-6-3-k2-18-7}
\subfigure[Frequency spectrum of $\theta_2$ for $k_1 = 6.3$, $k_2 = 18.7$]{
\includegraphics[width=\picwidth\textwidth]
{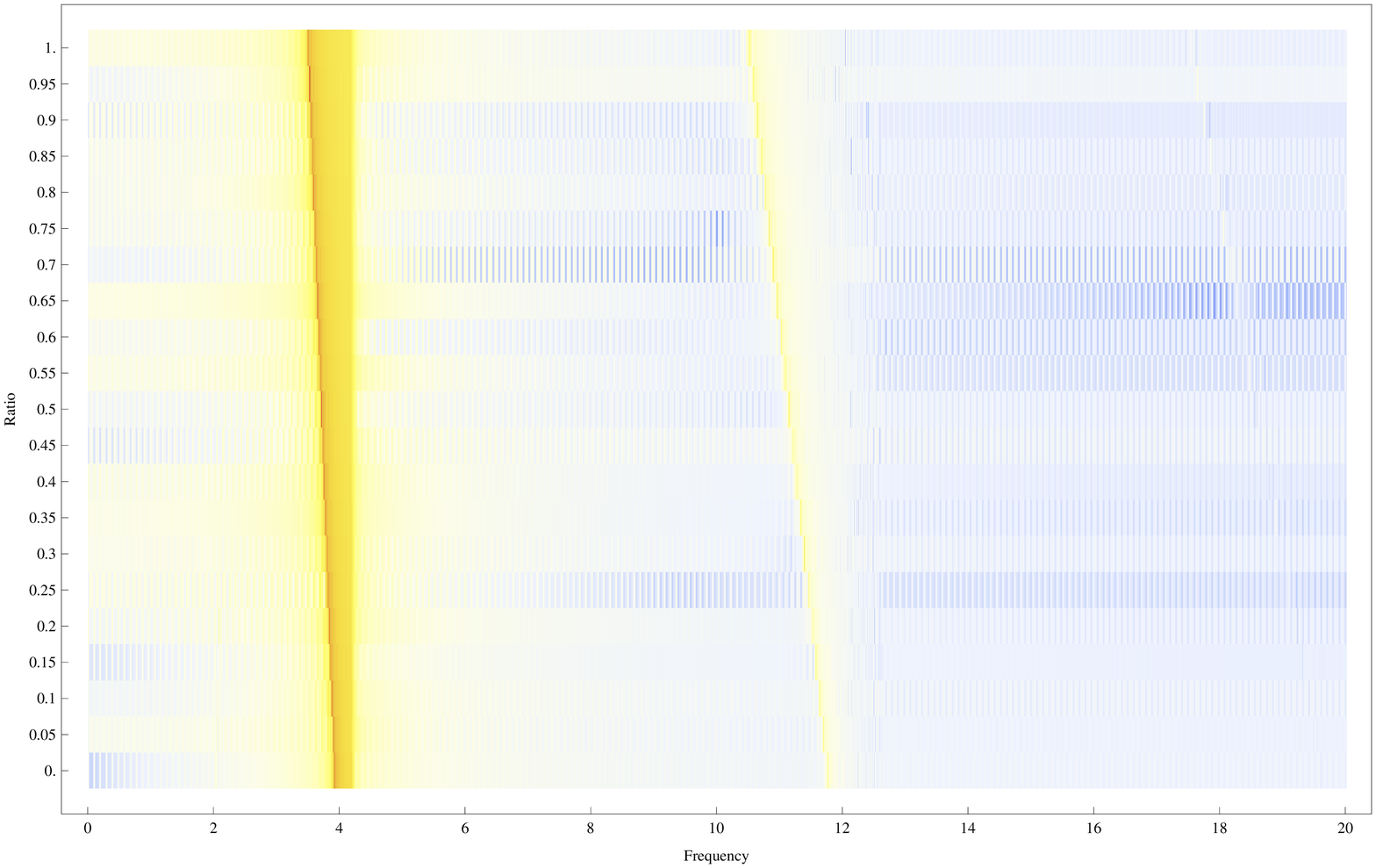}
}
\label{fig:FS-Theta2-k1-6-3-k2-18-7}
\caption{Frequency spectra for a range of $r$. In all the experiments $\theta_0 = 0.3$ and $\epsilon = 0.64$. The vertical axis shows the values for the parameter $r$ (ranging from 0.0 to 1.0), the horizontal axis shows the frequencies in the signal and the color shows the intensity of the corresponding frequency (blue shows lower intensity and red shows higher intensity)}
\label{fig:FS-for-Theta}
\end{figure*}


\subsection{Experiments}

\begin{figure}[htb]
\centering
\includegraphics[width=0.5\textwidth]{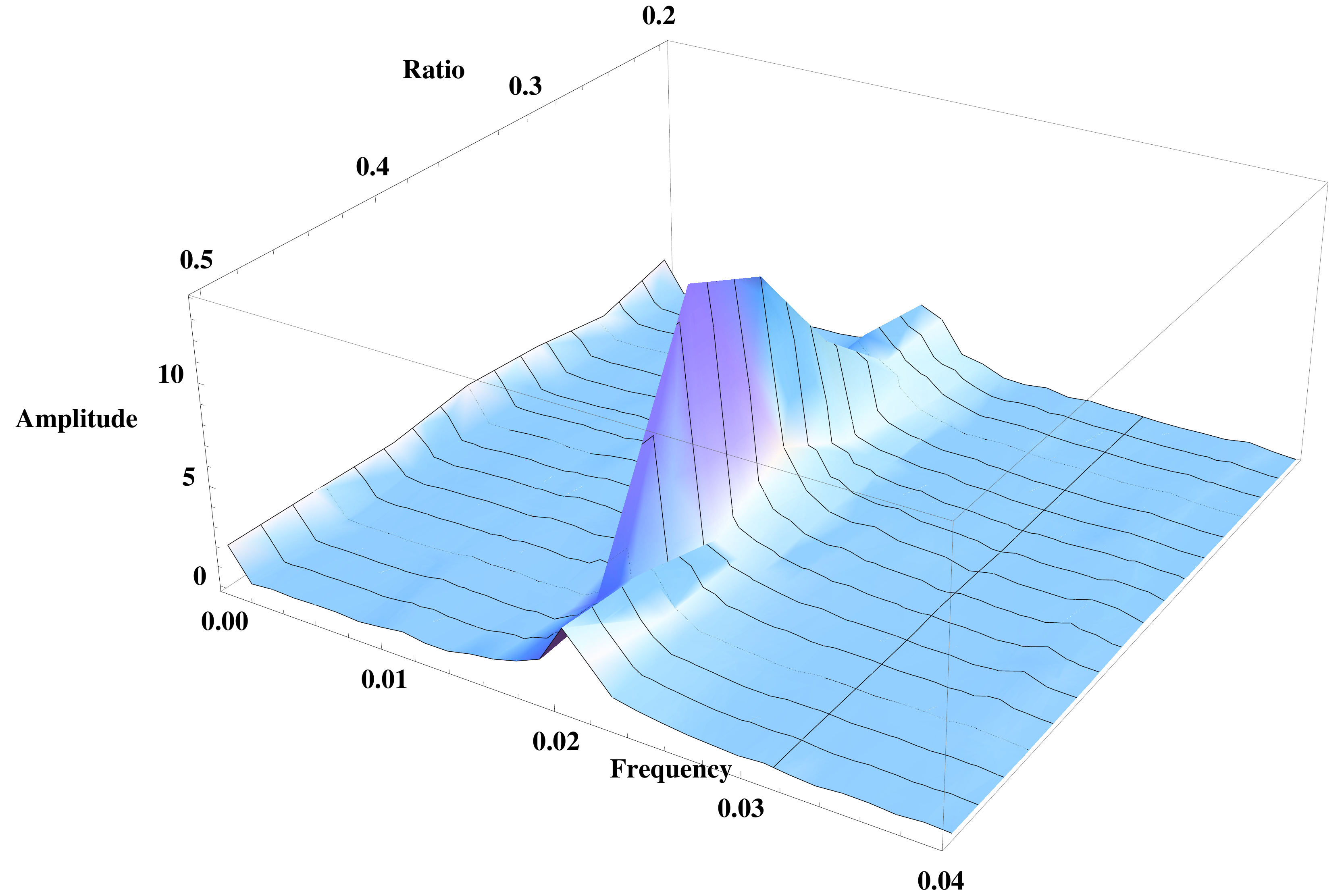}
\caption{Frequency spectra of $\theta_1 - \theta_2$ for different values of $r$ (ratio) in the experiments. In the experiments $r$ was fixed and the dynamics was observed for the specific value of $r$.}
\label{fig:20110525-1715-FreqSpec-Experiments}
\end{figure}

The motor in the experiments receives velocity as the desired input and this was the main reason we chose $y = f(\theta_1, \theta_2)$ format in the model to match the experimental limitations as shown in Eq. \ref{eq:ExperimentsMainEqTwoMetsOneActiveBase}\footnote{Please note that we have ignored the delay between setting the desired velocity and the time the move is applied to the cart.}.

\begin{subequations}
\label{eq:ExperimentsMainEqTwoMetsOneActiveBase}
\begin{align}
y &= \gamma_x (r \theta_1 + (1 - r) \theta_2) \\
\dot{z} &= \ddot{y}
\end{align}
\end{subequations}
in which $\gamma_x$ is an attenuating factor. Fig. \ref{fig:20110525-1715-FreqSpec-Experiments} shows the frequency spectra for experimental results described in Eq. \ref{eq:ExperimentsMainEqTwoMetsOneActiveBase}\footnote{To find out more accurate results, we need to run the experiments for much longer times. On the other hand, there are mechanical limitations (such as the limited metronomes wind). This is currently a limitation which leaves work for future research.}.

\section{Adaptation}


In this section we show how the system evolves when we introduce the possibility of adaptation.


\subsection{Two Oscillators on a Controlled Base with Adjusting Parameter}

In this case we also consider two metronomes placed on a moving rigid base, but we introduce a parameter for changing the behavior of the system. The metronomes oscillate freely while the movements of the base impact them. The movements of the base are a function of state variables with adjustments from a parameter. The equations of motion for this case are as shown in Eq. \ref{eq:MainEqTwoMetsOneActiveBaseWithParameter}

\begin{subequations}
\label{eq:MainEqTwoMetsOneActiveBaseWithParameter}
\begin{align}
\dot{x} &= y\\
\dot{y} &= z\\
\dot{z} &= p \theta_1 + (1 - p) \theta_2 \label{eq:MainEqTwoMetsOneActiveBaseWithParameterFifth} \\
\dot{f} &= LPF(\theta_1, \omega_1, \theta_2, \omega_2, z, p) \label{eq:MainEqTwoMetsOneActiveBaseWithParameterSixth} \\
\dot{p} &= f
\end{align}
\end{subequations}

\noindent in which $LPF(\theta_1, \omega_1, \theta_2, \omega_2, z, p)$ in Equation \ref{eq:MainEqTwoMetsOneActiveBaseWithParameterSixth} is a low pass filter for modifying the adjusting parameter ($p$). A sample function for $LPF$ is

\begin{equation}
\dot{f} = LPF(\theta_1, \omega_1, \theta_2, \omega_2, z, f) = f - \underbrace{(- 0.9 f + 0.9 \theta_1)}_{\text{Low pass filter}}
\end{equation}

\subsection{Low-Pass Filtered Feedback}

A low-pass filter is made to provide feedback. The feedback is used for adjusting the parameter which specifies the qualitative behavior of the system. The equivalent differential equation is written as

\begin{equation}
R_2 C \frac{dV_{out}}{dt} + V_{out} = - \frac{V_{in} R_2}{R_1}
\end{equation}

\noindent So, for our filter, we write the differential equation as follows
\begin{equation}
\frac{dV_{out}}{dt}  = \frac{1}{R_2 C} \left(- \frac{V_{in} R_2}{R_1} - V_{out}\right)
\end{equation}
\noindent In this filter the cutoff frequency (in hertz) is defined as

\begin{equation}
f_c = \frac{1}{2 \pi R_2 C}
\end{equation}

\noindent or equivalently (in radians per second):

\begin{equation}
\omega_c = \frac{1}{R_2 C}
\end{equation}

\noindent The gain in the passband is $− \frac{R_2}{R_1}$, and the stop-band  drops off at −6 dB per octave as it is a first-order filter. Hence, the differential equations of our system will change as follows
\begin{align}
\dot{p} = \alpha \left(\frac{1}{R_2 C}\right) \left(- p - \frac{\omega_1 \omega_2 R_2}{R_1}\right) \label{eq:MainEqTwoMetsOneActiveBaseWithParameterWithLowPassFilterSixth}
\end{align}
In the filter, the feedback is attenuated based on \cite{HW-SASAC-2009}. Also we have considered $R_1 = R_2 = R$, the gain of the filter is negative (set to be $-1$ for this case) and hence the negated value of $p$ is used in the differential equations.

\def \picwidth {0.22}
\begin{figure}
\centering
\subfigure[$p(0)=0.00$]{
\includegraphics[width=\picwidth\textwidth]
{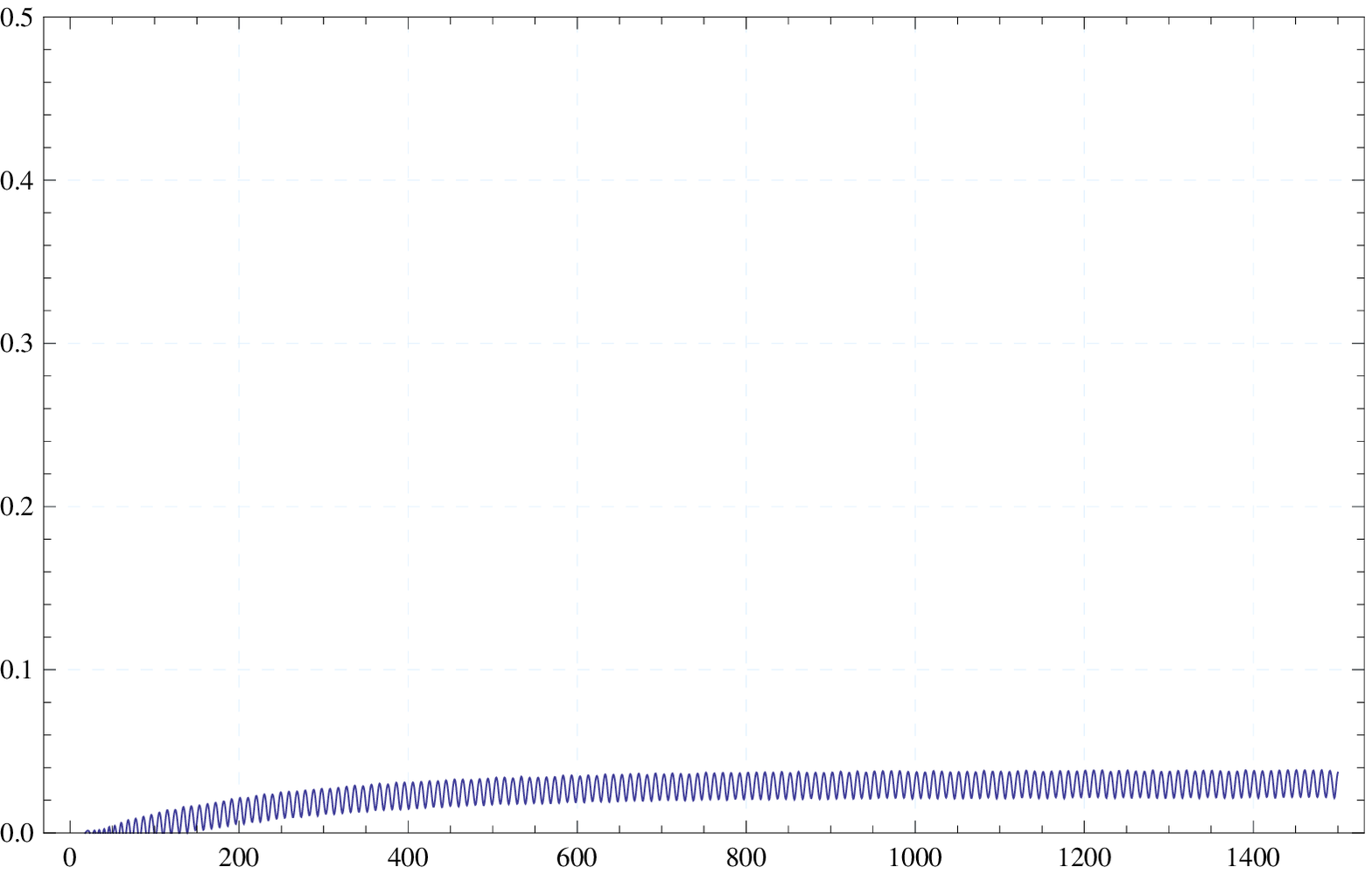}
}
\subfigure[$p(0)=0.15$]{
\includegraphics[width=\picwidth\textwidth]
{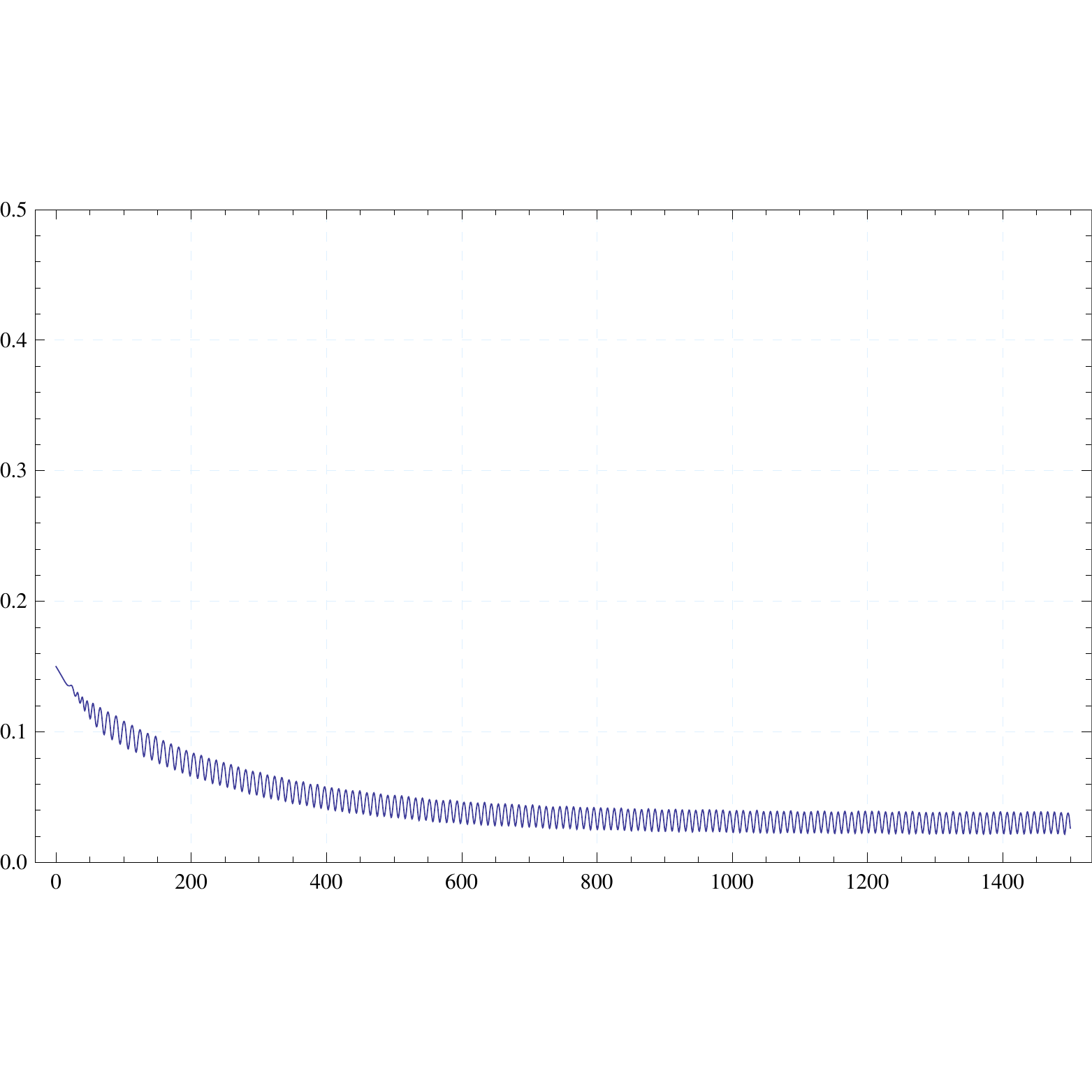}
}
\subfigure[$p(0)=0.30$]{
\includegraphics[width=\picwidth\textwidth]
{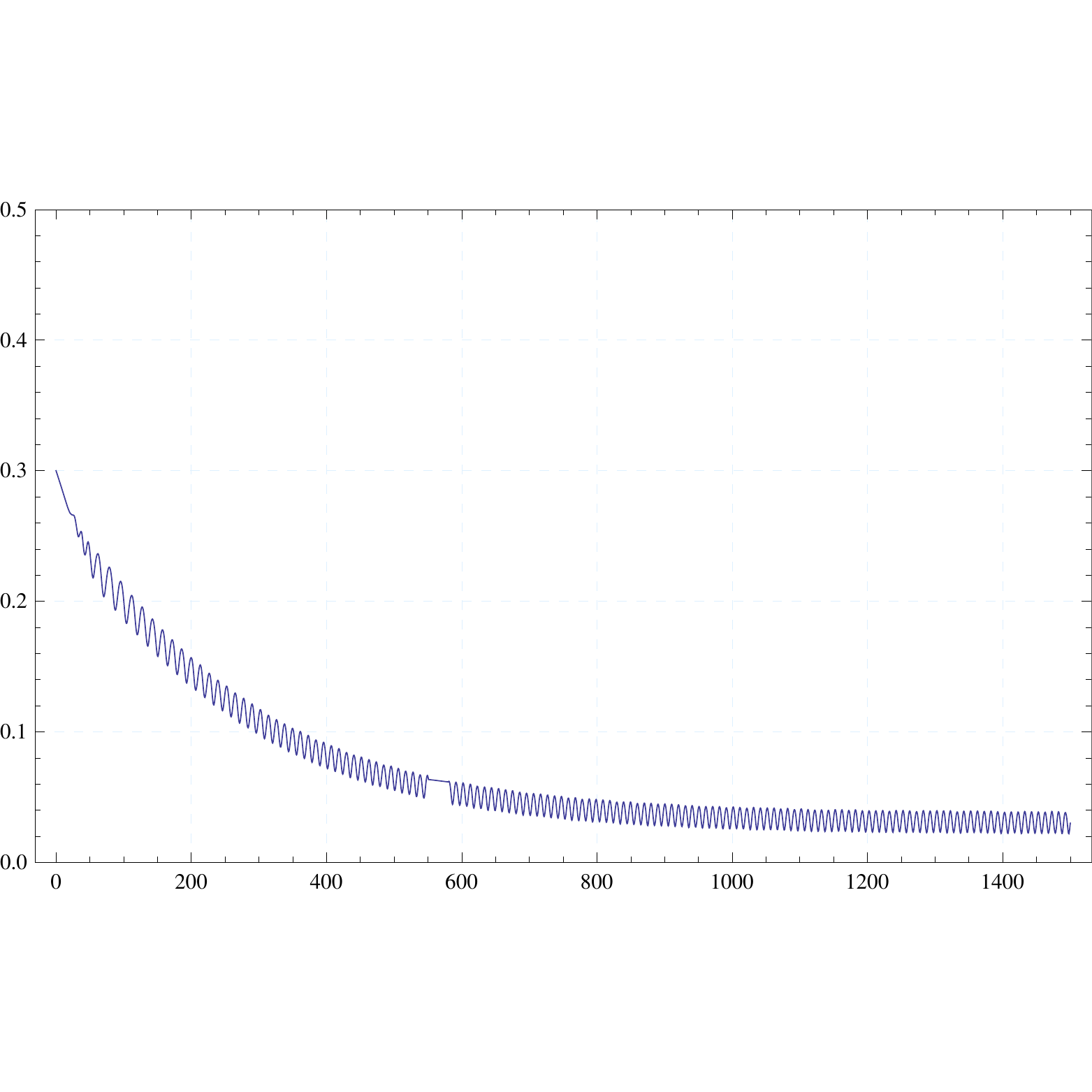}
}
\subfigure[$p(0)=0.45$]{
\includegraphics[width=\picwidth\textwidth]
{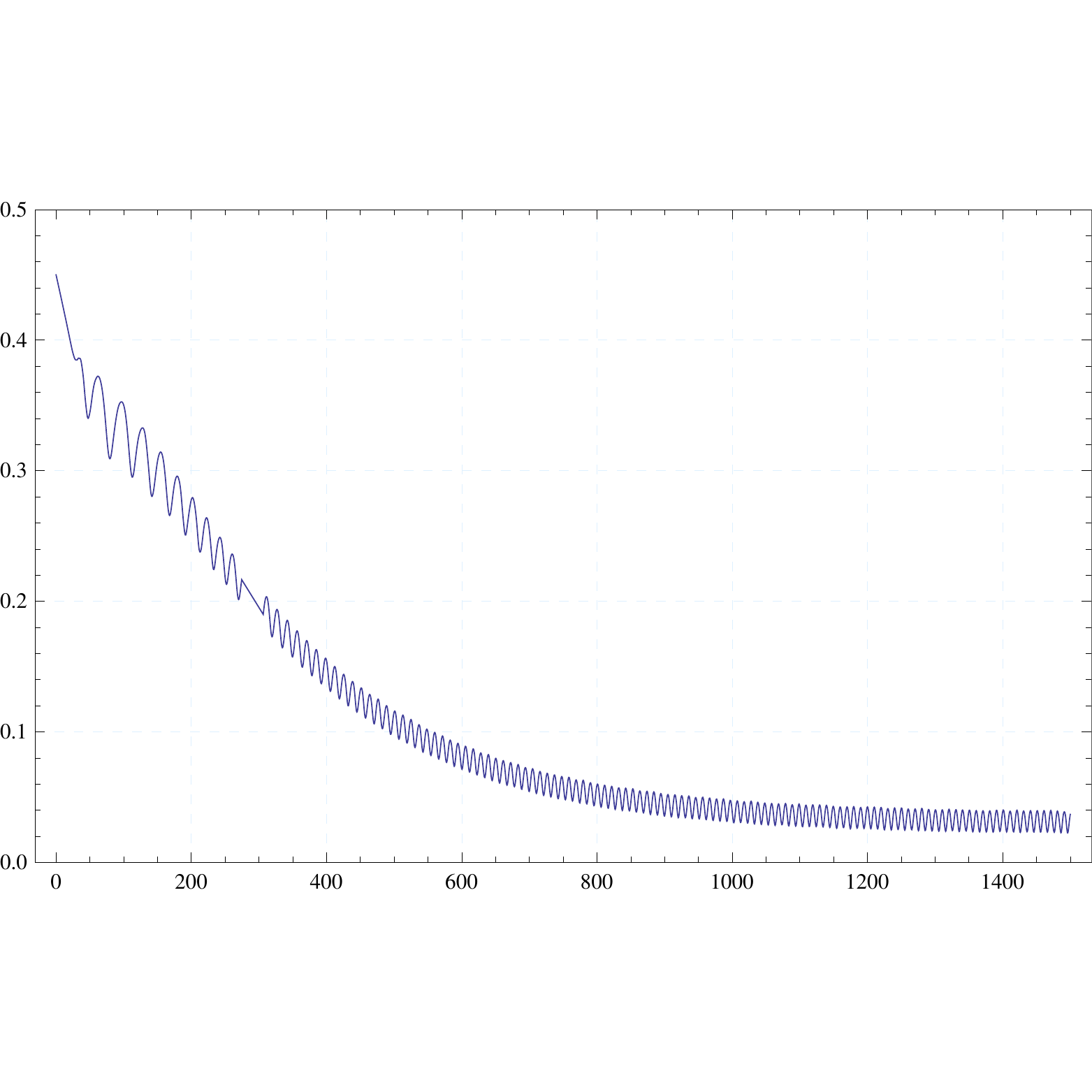}
}
\label{fig:20100925-1555-x9-initial-0-00-to-0-45}
\caption{Value of $p$ (the adjustable parameter) in time for initial values from $0.00$ to $0.45$.}
\end{figure}




\begin{figure}
\centering
\includegraphics[width=0.48\textwidth]{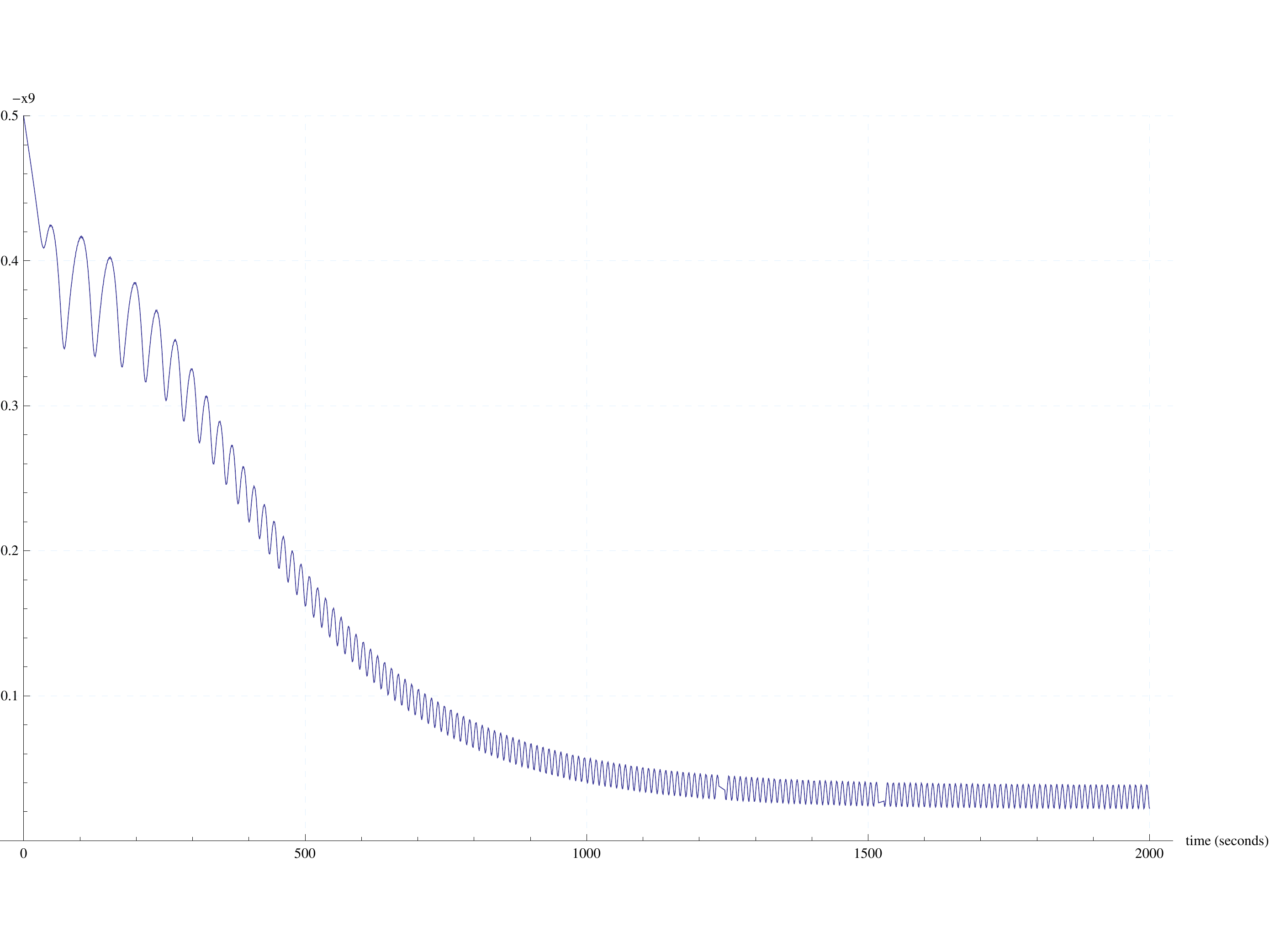}
\caption{Value of $p$ (the adjustable parameter) in time for initial value = $0.50$.}
\label{fig:20101021-1630-x9-initial-0-50-x9-without-frame}
\end{figure}

\subsection{Adaptation with the Experimental Limitations with Simple Function}

As it was mentioned earlier, the motor in charge of the cart receives velocity as the desired input. For this case we have considered 
\begin{equation}
y = \gamma_x (p \theta_1 + (1 - p) \theta_2)
\end{equation}
so
\begin{align}
z &= \dot{y}
\nonumber \\
  &= \gamma_x (\theta_1 \dot{p} - \theta_2 \dot{p} + p \dot{\theta_1} - (p - 1) \dot{\theta_2})
\end{align}
and
\begin{equation}
\begin{split}
\dot{z} &= \gamma_x (\theta_1 \ddot{p} - \theta_2 \ddot{p} + 2 \dot{p} (\dot{\theta_1} - \dot{\theta_2}) + p \ddot{\theta_1} - p \ddot{\theta_2} + \ddot{\theta_2})
\end{split}
\end{equation}


\subsubsection{$\dot{f} = \gamma_f (- f - (\theta_1 - \theta_2))$}

Assuming that
\begin{equation}
\begin{split}
\dot{f} &= \alpha \left(\frac{1}{R C}\right) (- f - (\theta_1 - \theta_2))
\\
        &= \gamma_f (- f - (\theta_1 - \theta_2))
\end{split}
\end{equation}
defines the dynamics of the parameter $f$, in which $\gamma_f = \alpha \left(\frac{1}{R C}\right)$, now we have
\begin{equation}
\begin{split}
\dot{p} &= \gamma_p f
\\
\ddot{p}&= \gamma_p \dot{f}
\\
        &= \gamma_p \gamma_f (- f - (\theta_1 - \theta_2))
\end{split}
\end{equation}
so, we re-write the relation for $\dot{z}$ as
\begin{equation}
\begin{split}
\dot{z} &= \gamma_x (\theta_1 (\gamma_p \gamma_f (- f - (\theta_1 - \theta_2)))
\\
&- \theta_2 (\gamma_p \gamma_f (- f - (\theta_1 - \theta_2)))
\\
&+ 2 \gamma_p f (\omega_1 - \omega_2) + p \dot{\omega}_1 + (1 - p) \dot{\omega}_2)
\end{split}
\end{equation}
replacing right hand sides of $\dot{\omega}$ yields
\begin{equation}
\begin{split}
\dot{z} &= \gamma_x (\theta_1 (\gamma_p \gamma_f (- f - (\theta_1 - \theta_2)))
\\
&- \theta_2 (\gamma_p \gamma_f (- f - (\theta_1 - \theta_2))) + 2 \gamma_p f (\omega_1 - \omega_2)
\\
&+ p (- k_1 \sin{(\theta_1)} - \epsilon \left[\left(\frac{\theta_1}{\theta_0}\right)^2 - 1\right] \omega_1 - \frac{k_1}{g} \cos{(\theta_1)} z)
\\
&+ (1 - p) (- k_2 \sin{(\theta_2)} - \epsilon \left[\left(\frac{\theta_2}{\theta_0}\right)^2 - 1\right] \omega_2 - \frac{k_2}{g} \cos{(\theta_2)} z))
\end{split}
\end{equation}
\begin{figure}[]
\centering
\includegraphics[width=0.45\textwidth]{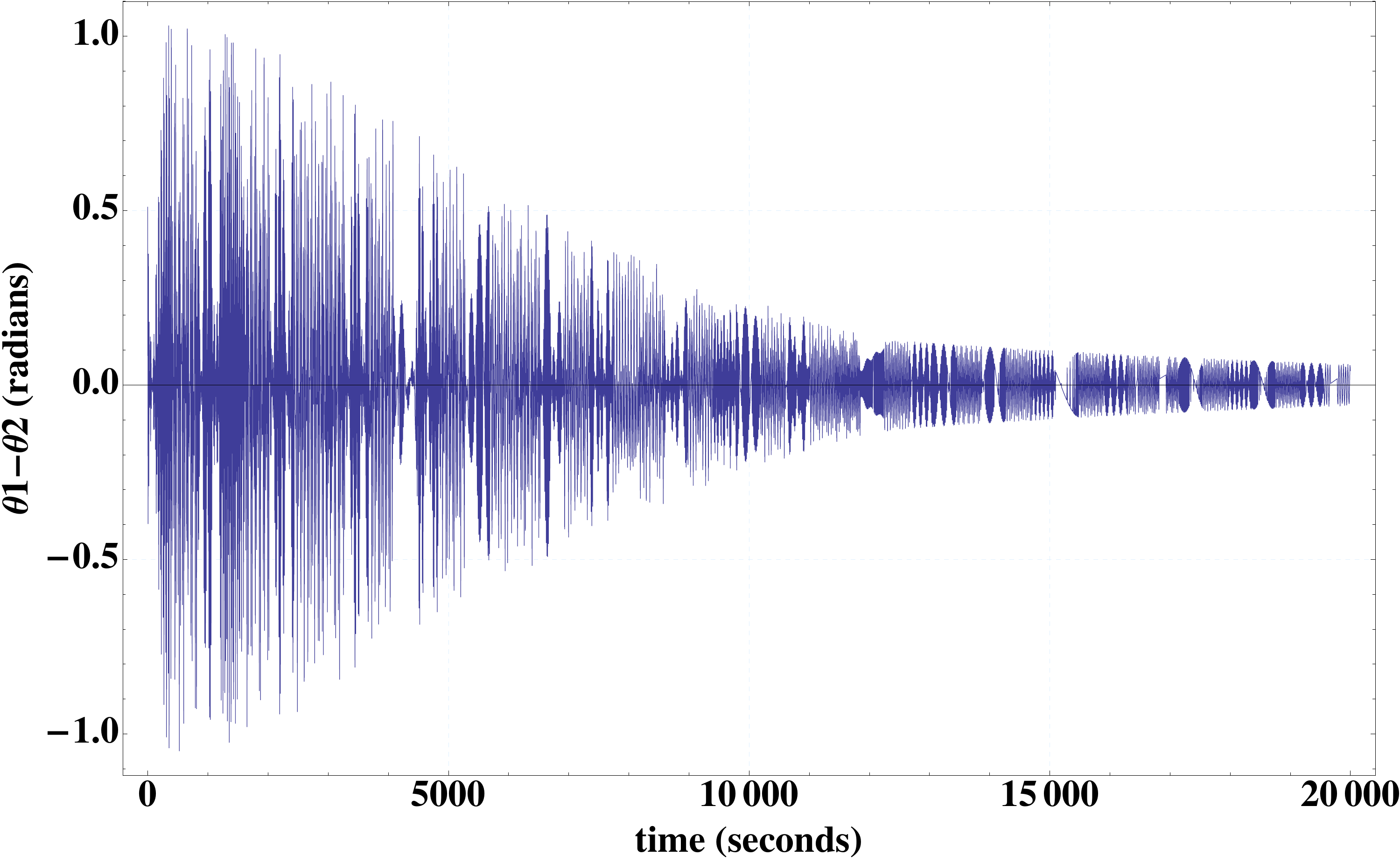}
\caption{$\theta_1 - \theta_2$ when the parameter adapts}
\label{fig:20110827-1938-theta1-minus-theta2-k1-6-4-k2-12-8}
\end{figure}
\begin{figure}[]
\centering
\includegraphics[width=0.45\textwidth]{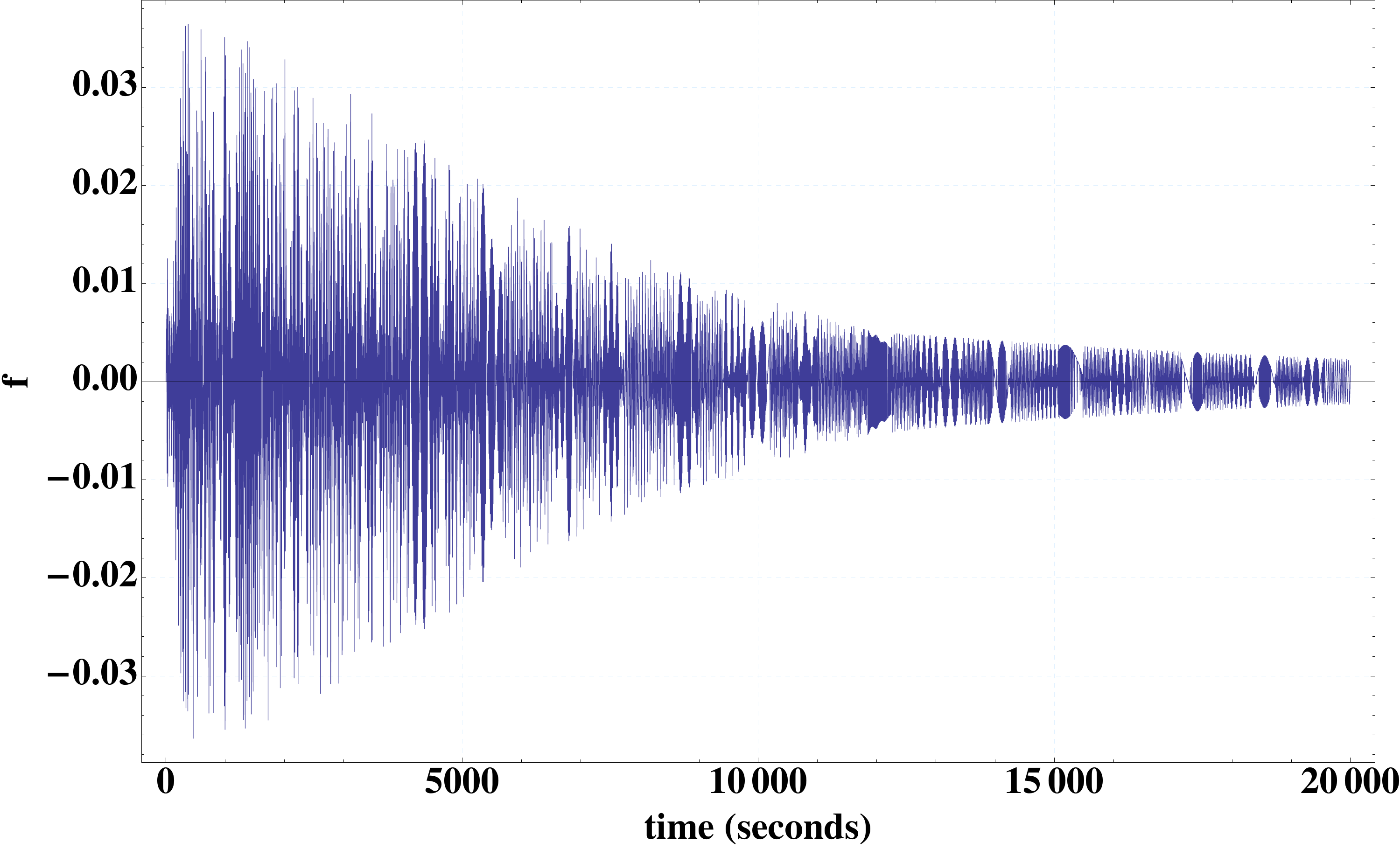}
\caption{$f$ when the parameter adapts}
\label{fig:20110827-1938-f-k1-6-4-k2-12-8}
\end{figure}
\begin{figure}[]
\centering
\includegraphics[width=0.45\textwidth]{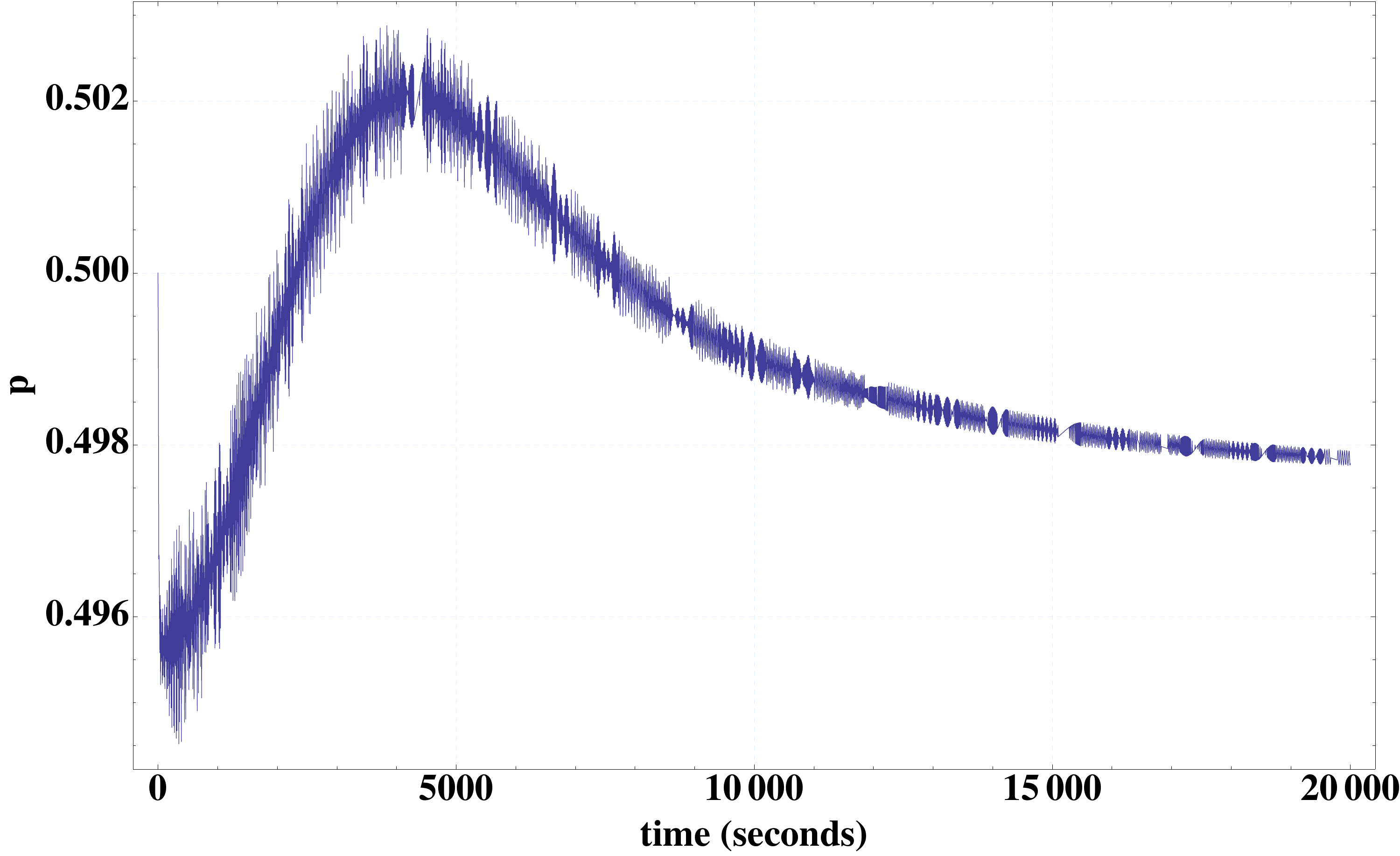}
\caption{$p$ when the parameter adapts}
\label{fig:20110827-1938-p-k1-6-4-k2-12-8.eps}
\end{figure}


\subsection{Adaptation in Experiments}

In an experiment the parameter was set to be adjusted via a low-pass filtered feedback of $\theta_1 \theta_2$ state variables. The change of this adjusting parameter in time for two different tests are shown in Fig. \ref{fig:adjusting-parameter-in-experiment}. We can find a sense of the meaning of these values for the parameter when looking at the frequency spectrum for experiments shown in Fig. \ref{fig:20110525-1715-FreqSpec-Experiments}.

\def \picwidth {0.45}
\begin{figure}[htb]
\centering
\subfigure{
\includegraphics[width=\picwidth\textwidth]
{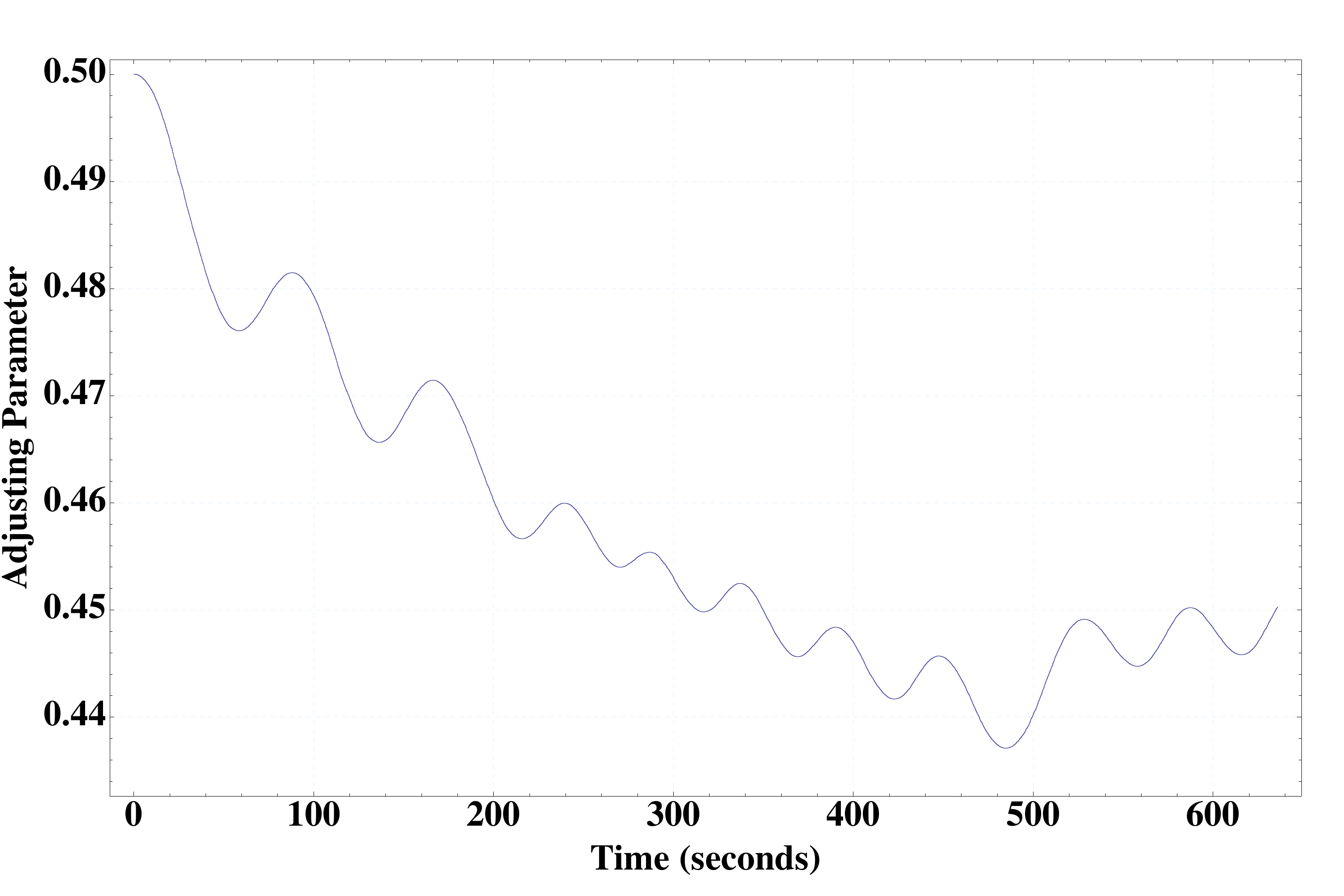}
}
\subfigure{
\includegraphics[width=\picwidth\textwidth]
{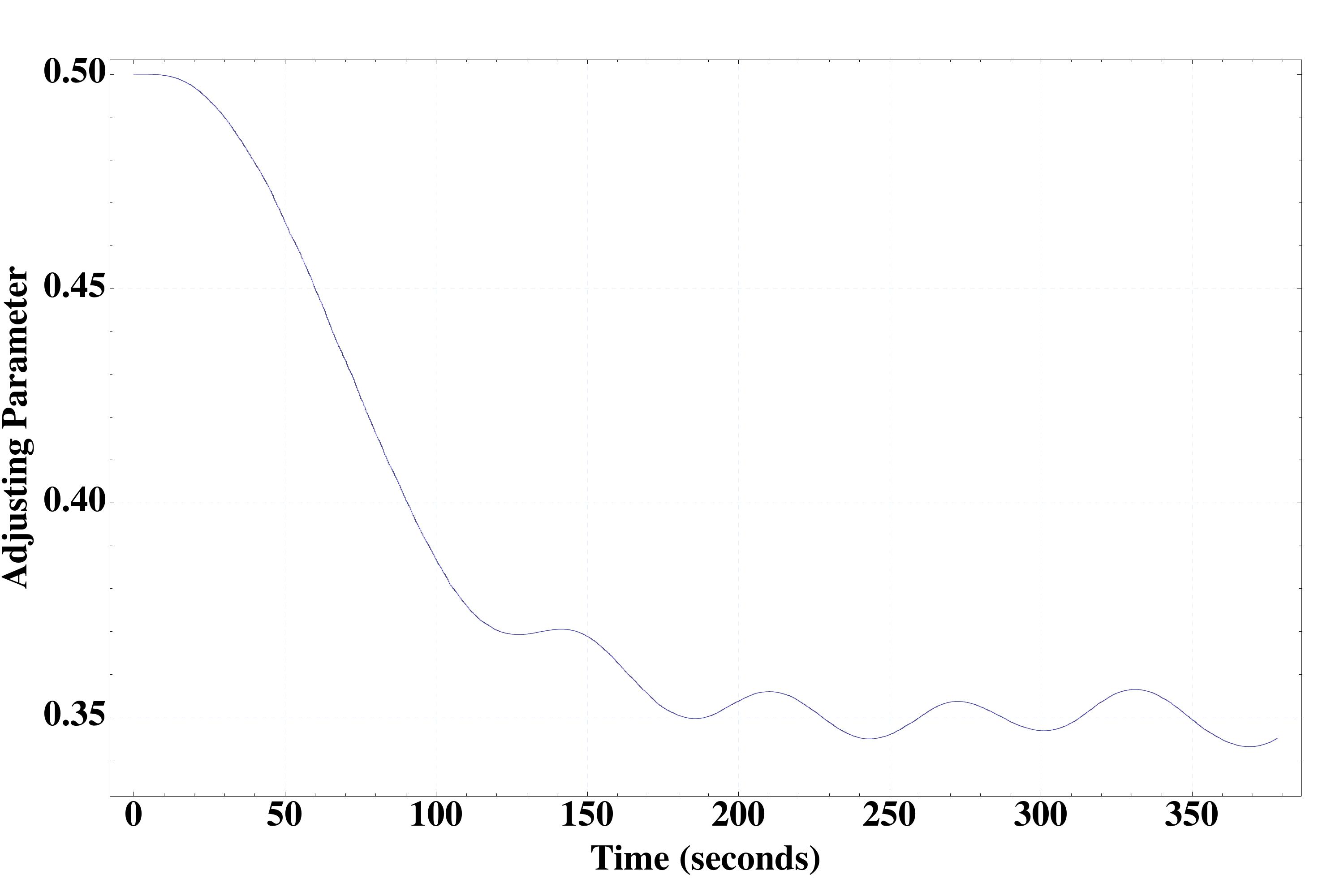}
}
\caption{Change of the adjusting parameter in two different experimental runs (the values are different because of different initial conditions in the experiments)}
\label{fig:adjusting-parameter-in-experiment}
\end{figure}

\section{Conclusions and Future Work}


In this section we have conclusions of this research and will provide some recommendations for future work.


\subsection{Conclusions}

In this article we studied the dynamics of two oscillators that are mechanically coupled through a rigid base. The base was controlled with a motor and the movements of the base affected the metronome rods. We showed that how the functions controlling the base, determined the dynamics of the system. The function in charge of moving the base had parameters and we introduced adaptation in the values of these parameters. We showed that how the parameters changed and how was the dynamics of the system after change of the parameters. We showed that when we introduce adaptation of the adjusting parameter, it settles to the direction that the dynamics of the system is in a regime between order and chaos\footnote{In other words, the dynamics of the system goes towards a region between lowest and highest number of frequencies in the frequency spectrum.}.


\subsection{Future Work}
\label{sec:FutureWork}

There are some improvements that are suggested for the future work of this research. For analysis and numerical simulation, we considered an ideal model of the system in which there are not any delays or random terms. An area for future work would be to include these terms in the model and investigate the system. The similarity to the experimental results would show the accuracy of the model.

Using other alternatives for detecting the phases of the oscillators is also a possible improvement for the system. This might include using sound detecting devices \footnote{The metronomes make a tick sound when the rod reaches the far left or far right.} and encoders for detecting the angles of the metronomes \footnote{Reading the values of the angles by using mechanical devices is reasonable if the sensory device does not interfere with the dynamics of the system. One of the main reasons that we had more tendency towards vision is that it does not interfere with the oscillations of the metronome rods and their dynamics.}. Other devices for sensing can be used simultaneously so that the accuracy of the data is increased.

The oscillators are mechanically coupled in this system. A possible direction of future research would be considering a case in which the coupling involves delay. This can be realized when the metronomes are positioned on two independent carts and each of the carts is capable of moving independently. The same can be done for when there are random terms in the coupling between oscillators.

Another possible area for extending the current work would be to introduce more oscillators. The number of oscillators and their positioning besides how they are connected to each other are some parameters of the system to be determined. A higher number of oscillators will certainly introduce more complexity to the system and we expect to see more complex (and maybe richer) dynamics. We would also recommend using other types of oscillators that are not necessarily mechanical (for instance electrical circuits). A similar system can be replicated with other oscillators and the evolution of the system can be studied.

In this research we showed some cases and considered the evolution of the system with the relevant parameters and initial conditions. A possible area to be investigated is studying the space of parameters and the evolution of the system for different regions, i.e. what is the role of initial conditions and parameters of the system (such as dynamics of the cart) on the evolution of the system.


\section*{Acknowledgements}
I would like to appreciate Dr. Tam\'as Kalm\'ar-Nagy for helping me with this research. This work would not have been possible without our long discussions, his constructive feedbacks and his continuous support throughout the last three years.  I would also like to thank Ali-Akbar Agha-Mohammadi for long discussions and his constructive comments, Tracey Neal Thompson for helping with experiments, Dr. Kevin Hernandez for his help with simulations and Amir Salimi for his help with analysis of the signals.


\bibliographystyle{unsrt}
\bibliography{RoozbehBibliography}


\clearpage
\renewcommand{\theequation}{A-\arabic{equation}}
\setcounter{equation}{0}  
\section*{APPENDIX}
\appendix
\section{Experimental Settings}
\label{sec-Experiments}


The experimental settings were used to implement the proposed model and validate the numerical results for experiments. The description and details of these settings are provided in this section.


\subsection{Description}

\begin{figure}[htb]
 \centering
 \includegraphics[width=0.45\textwidth]{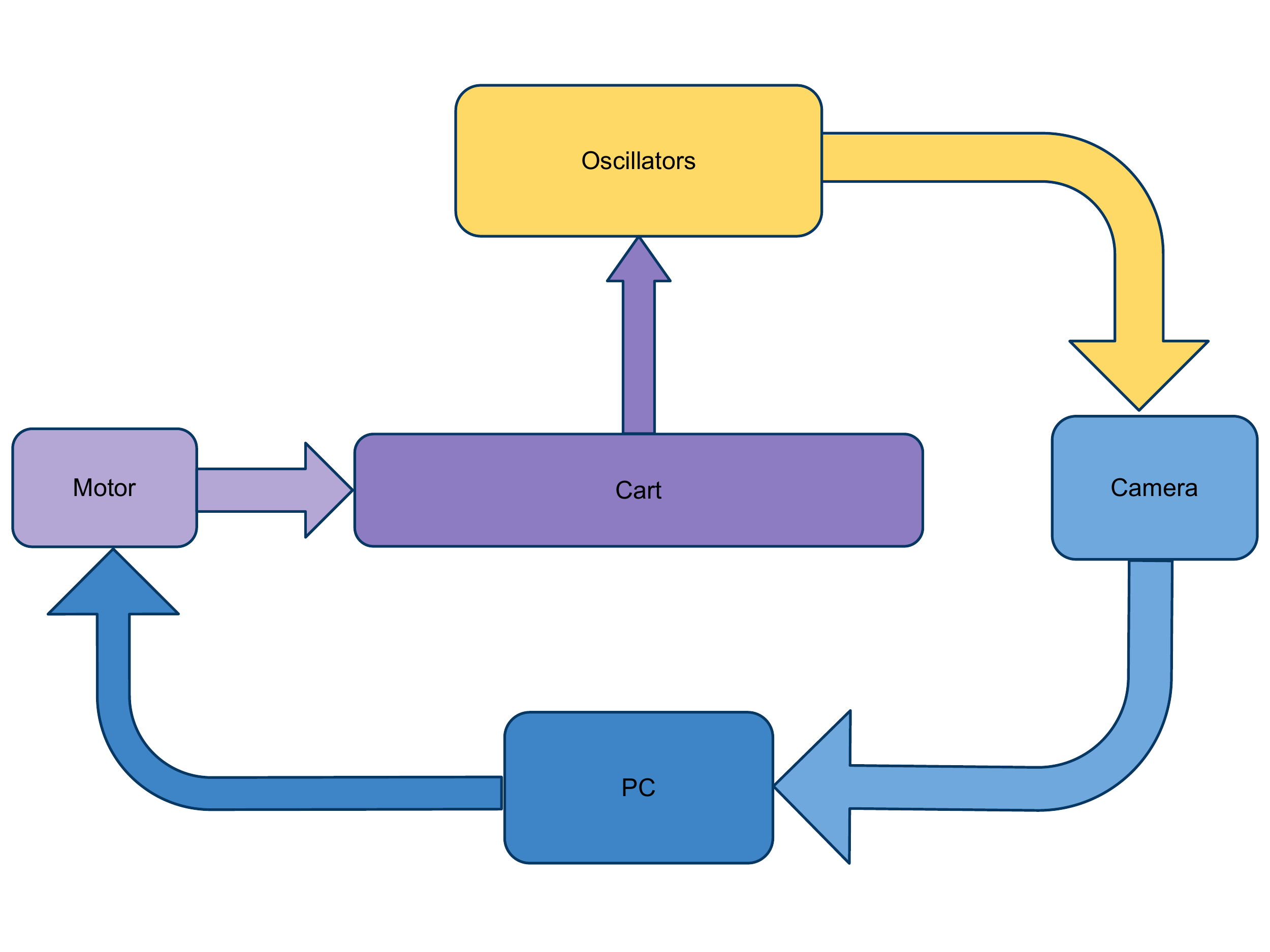}
 \caption{The schematic of the system}
 \label{fig:20100916-1844-LifeGeneralDiagram}
\end{figure}

We use mechanical wind-up metronomes as oscillators that are mechanically coupled through a moving base. The metronomes are Wittner's Super-Mini-Taktell (Series 880) and are claimed to be the world's smallest pendulum metronomes. Another part of the system is in charge of reading the angle values of metronome rods in time. A schematic of the system is shown in Fig. \ref{fig:20100916-1844-LifeGeneralDiagram} and some images of the experimental settings are shown in Fig. \ref{fig:20100920-1650}.

\def \picwidth {0.23}
\begin{figure*}
\centering
\subfigure[Stickers used for image processing]{
\includegraphics[width=\picwidth\textwidth]
{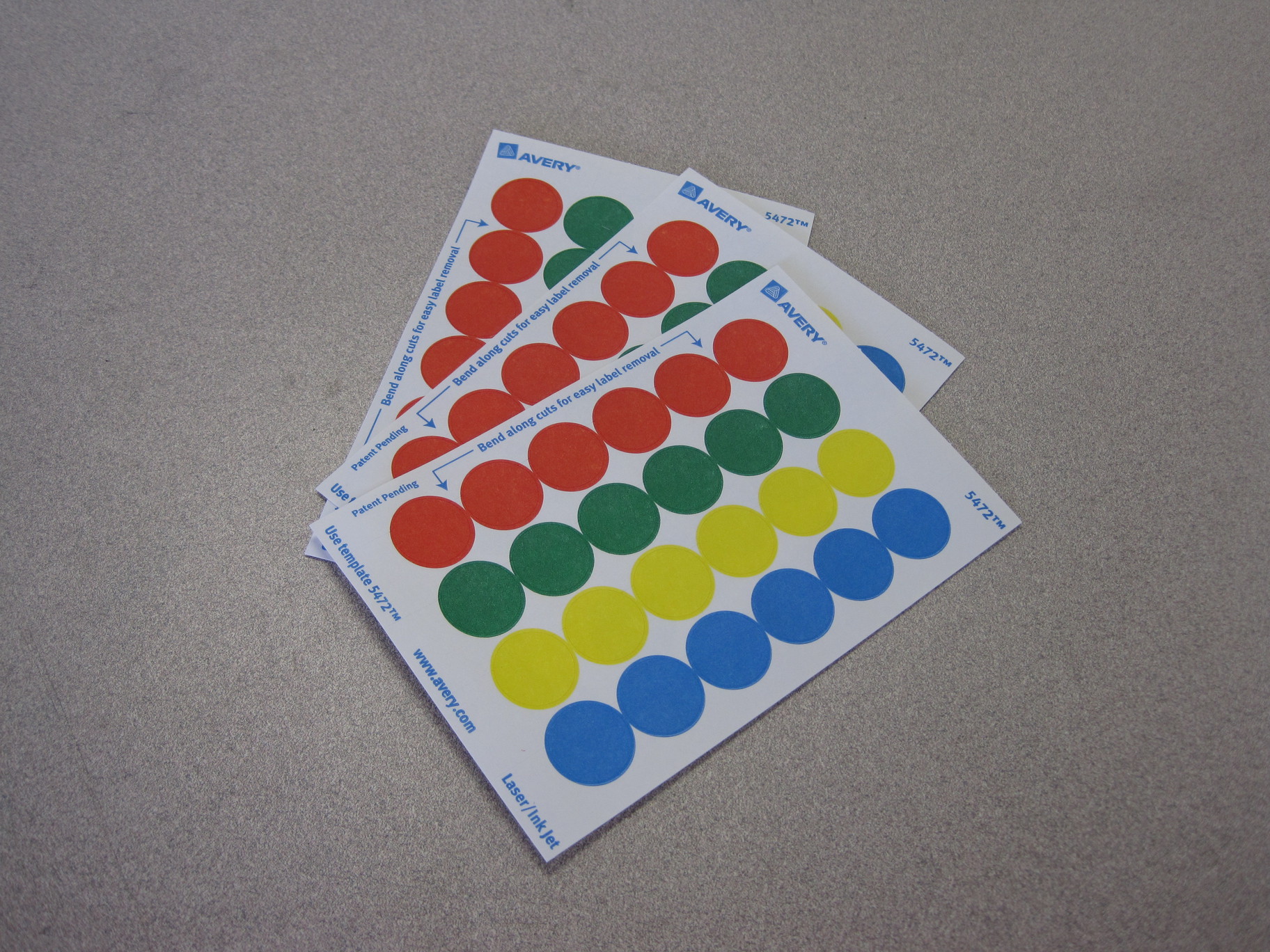}
}
\subfigure[Two metronomes on a freely moving cart]{
\includegraphics[width=\picwidth\textwidth]
{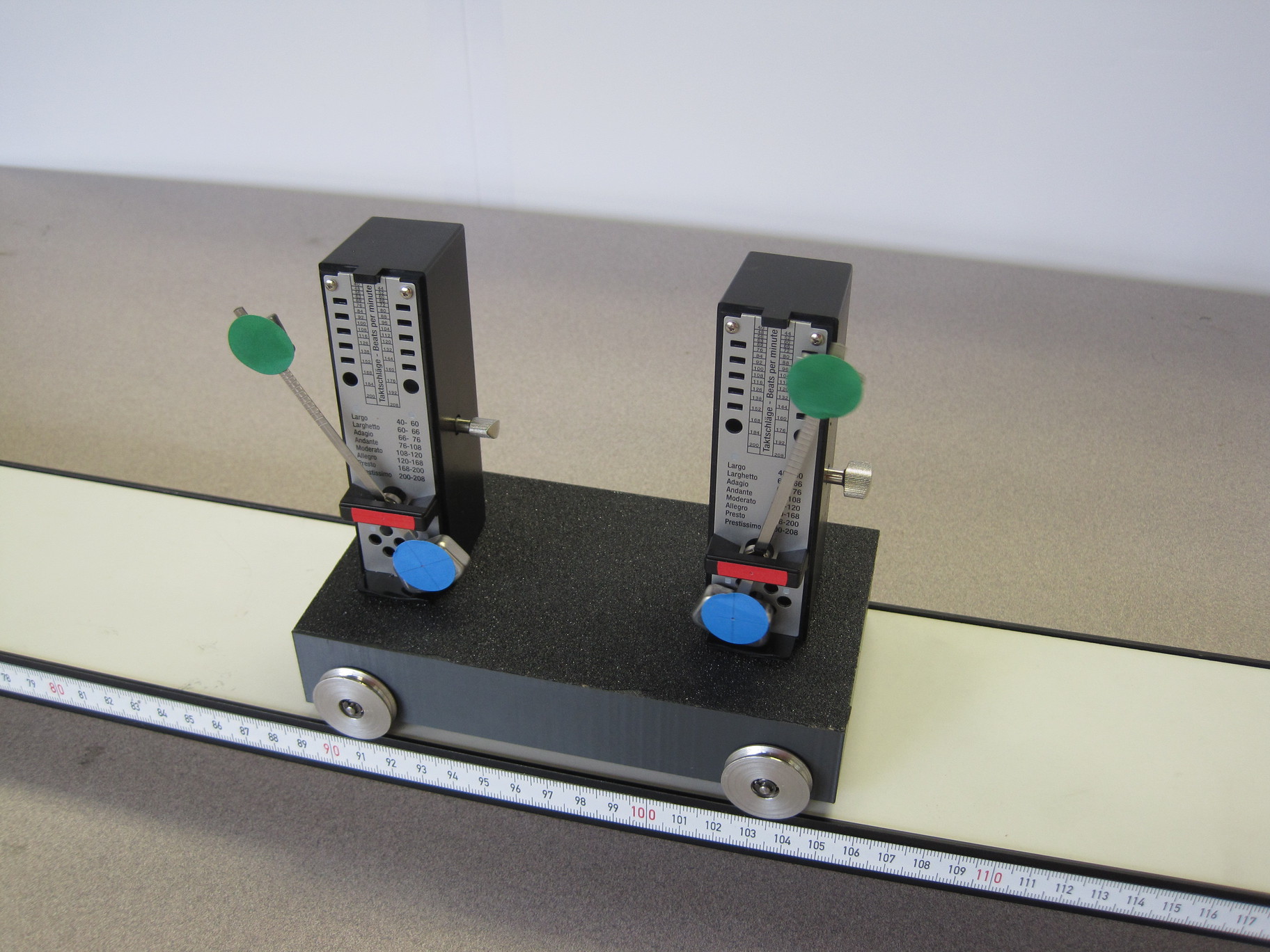}
}
\subfigure[Servo motor and the controller for moving the cart]{
\includegraphics[width=\picwidth\textwidth]
{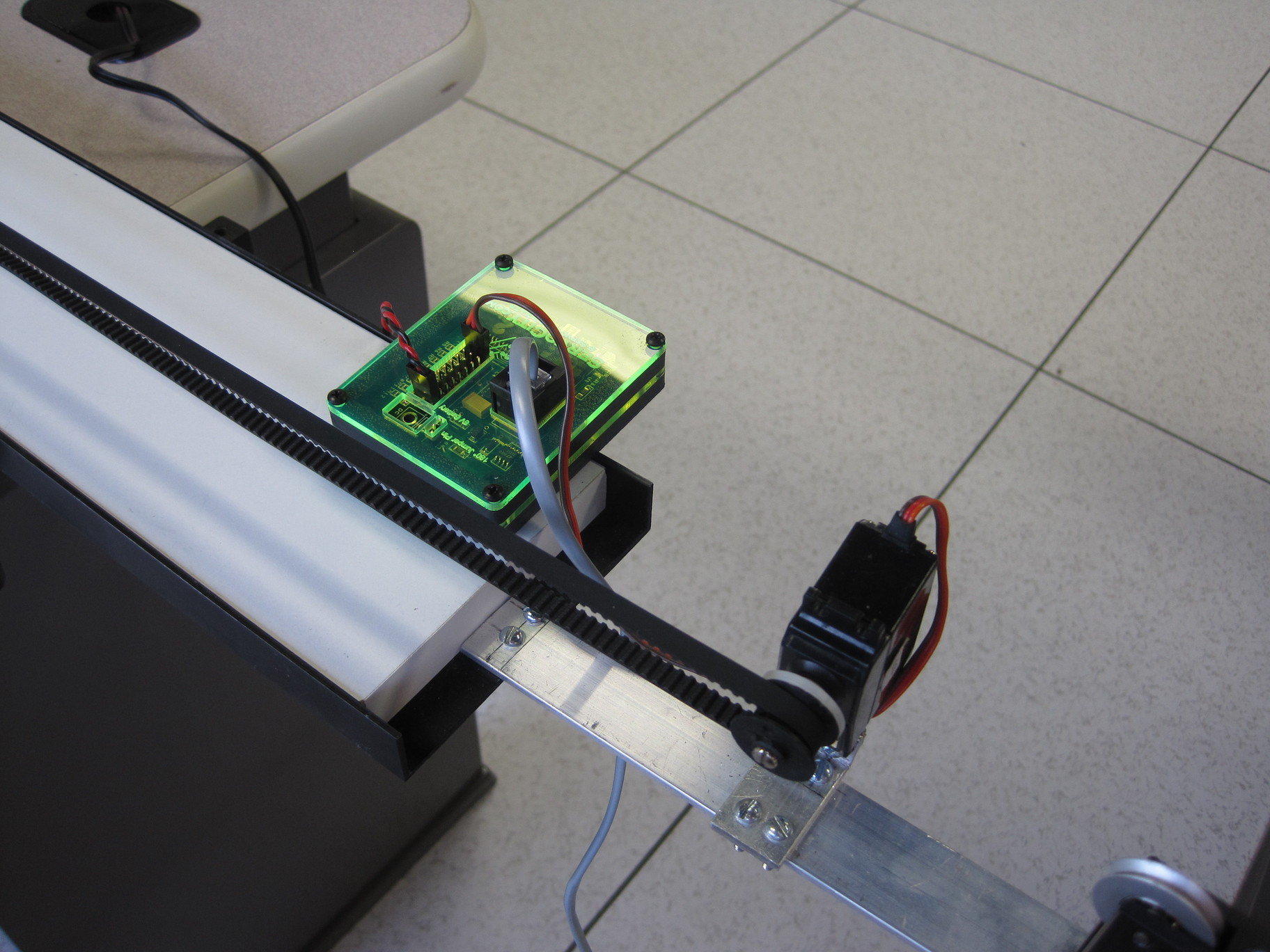}
}
\subfigure[A general view of the whole settings including the camera, the track, the carts, the motor and the pc]{
\includegraphics[width=\picwidth\textwidth]
{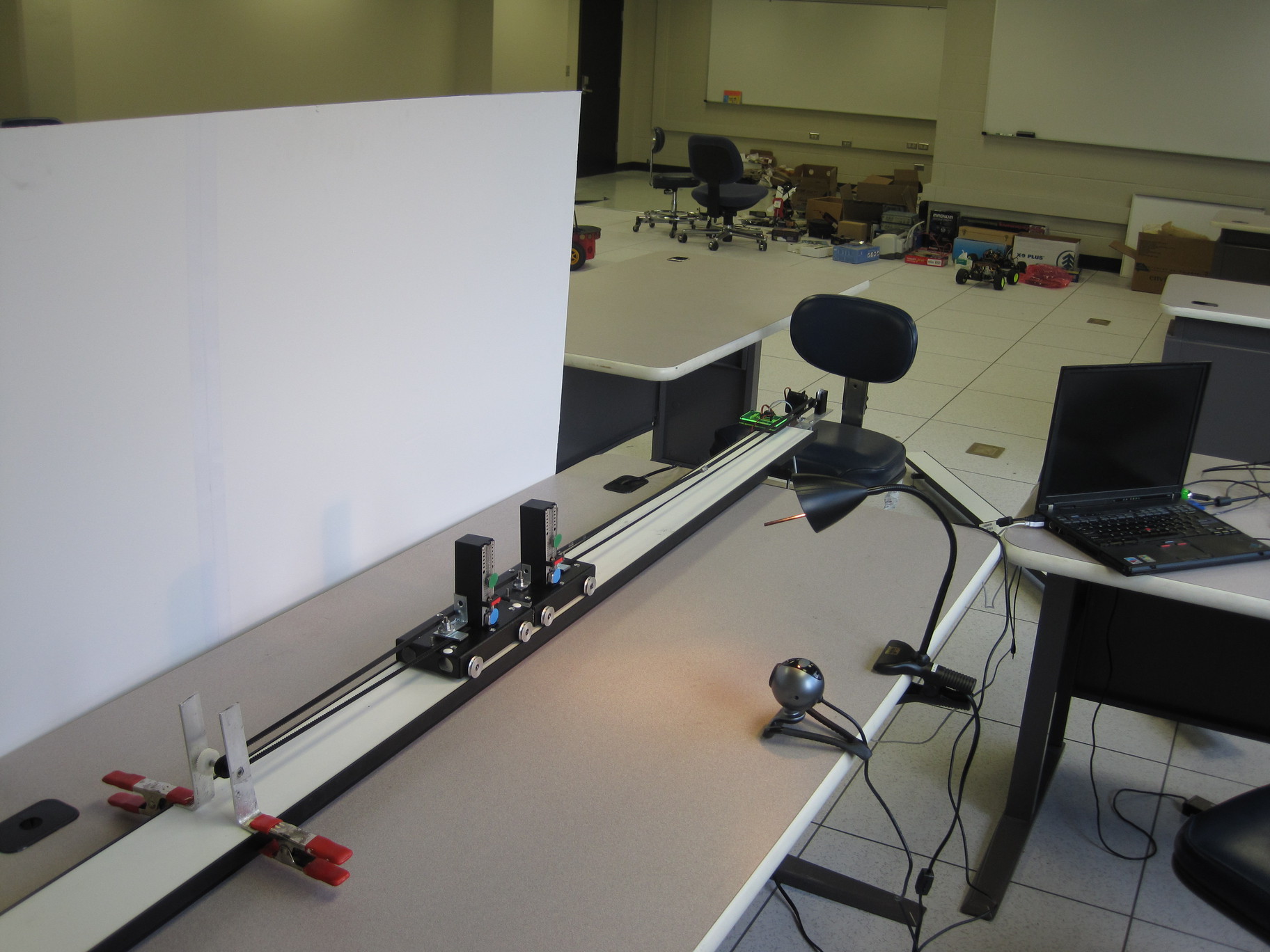}
}
\caption{Experimental Settings}
\label{fig:20100920-1650}
\end{figure*}


\subsection{Settings}

One of the main factors we had in mind was that to build a robust and reliable setting that is easily available and possible to use. We needed to read the values of angles at different times and one of the first candidates was using encoders for reading the values. As the encoders might have affected the mechanics of the metronomes and hence might have affected the dynamics, we preferred to choose an approach for reading the values of angles that does not affect the dynamics of the system.

We selected a vision and an image processing approach so that we track the rod; for instance, the conservation of mechanical energy using video analysis has been investigated in \cite{b-icmeuvafc-2010}. We used simple colored stickers so that the vision system can use the color for tracking the position. The stickers are circles for which the radius is 9.4 mm and the choice of colors are red, green, yellow and blue. Regarding the position of the stickers, the center of each sticker was marked and a tiny hole was made in the center. This hole was used to align the position of the sticker on the center of the rod.


\subsection{Image Processing}

\begin{figure}[htb]
 \centering
 \includegraphics[width=0.45\textwidth]{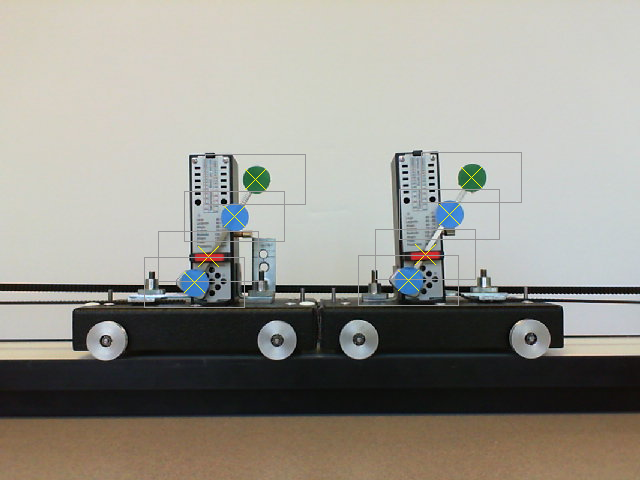}
 \caption{Two metronomes on two connected carts. The crosses show the center of mass of the detected stickers. Also there are rectangles around each of the stickers which show the area to be searched at each step.}
 \label{fig:8110000}
\end{figure}

The main image processing library that is used in this system is OpenCV \footnote{OpenCV is a computer vision library originally developed by Intel. It is free for use under the open source BSD license. The library is cross-platform and it focuses mainly on real-time image processing. It is available at http://opencv.willowgarage.com/}. A simple USB webcam is used to capture the image and the capture frame rate is around 20 to 40 frames per second \footnote{This is a rough estimate and is extracted based on experiments in normal conditions. It is changed according to the settings of the program, hardware specifications and the camera used for image processing.}. We do not use a pre-recorded stream for processing so that we can have real-time control on the system. In order to track the objects in real-time, an algorithm is in charge of looking for proper pixels in a certain window around the object that is being tracked. With enhanced settings, we used more than one color sticker for each metronome rod so that we increase the accuracy of angle detection (color image processing has been used in some applications such as \cite{AT-CIPNTRT-1990}). A sample of the perceived image is shown in Fig. \ref{fig:8110000}.

%
%
%
%
%


\subsubsection{Finding the Angle}

When using more than two stickers on the rod, we need to find the least squares regression line. We use the following formula for finding the angle of the metronome rod

\begin{equation}
\theta = \tan^{-1}{ \left( \frac{n \sum\limits_{i = 1}^{n} x_i y_i - \sum\limits_{i = 1}^{n} x_i \sum\limits_{i = 1}^{n} y_i}{n \sum\limits_{i = 1}^{n} {x_i}^2 - \left( \sum\limits_{i = 1}^{n} x_i \right)^2} \right) }
\end{equation}

\noindent in which $n$ is the number of stickers and $x_i$ and $y_i$ are the $x$ and $y$ of the $i\textit{th}$ sticker in pixels.


\subsubsection{Validating the Results}

The image processing module might occasionally lose track of stickers. To prevent the consequences of this problem and to validate the results, correlation of the positions of the detected stickers are calculated as below

\begin{align}
r_{xy} = \frac{n\sum x_iy_i-\sum x_i\sum y_i}
{\sqrt{n\sum x_i^2-(\sum x_i)^2}~\sqrt{n\sum y_i^2-(\sum y_i)^2}}
\end{align}

\noindent in which $n$ is the number of stickers and $x_i$ and $y_i$ are the $x$ and $y$ of the $i\textit{th}$ sticker in pixels. If correlation is more than a certain threshold, the sample is accepted, used for updating the values and stored. If the correlation is smaller than the threshold, then there might be one possibility: The positions of the detected stickers are too close to a vertical line (which reduces the correlation). In that case, the detected points are virtually rotated $\pi/4$ radians and the correlation is calculated again. If the correlation is more than the threshold, it means that the detected points have been sufficiently close to a line and the line has been vertical on the first round. If correlation is less than the threshold even in the second round, then it shows error in detecting the stickers and the sample is discarded.


\subsection{Model of Oscillators}

The equation that governs the motion of a single metronome is described as \cite{P-SM-2002,UMP-SCDCMM-2009}
\begin{equation}
\frac{d^2 \theta}{d t^2} + \underbrace{\frac{m r_{c.m.} g}{I}}_{k} \sin{(\theta)} + \epsilon \left[\left(\frac{\theta}{\theta_0}\right)^2 - 1\right] \frac{d \theta}{d t} = 0
\end{equation}
which is rewritten as
\begin{subequations}
\begin{align}
\dot{\theta} &= \omega \\
\dot{\omega} &= - k \sin{(\theta)} - \epsilon \left[\left(\frac{\theta}{\theta_0}\right)^2 - 1\right] \omega
\end{align}
\end{subequations}
\noindent in which $\theta$ is the phase of the metronome (angle made with the vertical line), $I$ is the moment of inertia of the pendulum, $m$ is the mass of the pendulum, $r_{c.m.}$ is the distance of the pendulum's center of mass from the pivot point, $\epsilon$ specifies the effect of escapement and damping, $\theta_0$ is the van der Pol term and $g$ is the acceleration of the gravity. $k = \frac{m r_{c.m.} g}{I}$ determines the frequency of oscillation. The damping function $D(\theta)$ is defined as \cite{UMP-SCDCMM-2009}
\begin{equation}
D(\theta) = \left(\frac{\theta}{\theta_0}\right)^2 - 1
\end{equation}
and with small $\epsilon$, this term produces small oscillations with an amplitude of approximately $2 \theta_{0}$ in an isolated oscillator \cite{P-SM-2002}.

\begin{figure*}[htb]
 \centering
 \includegraphics[width=0.9\textwidth]{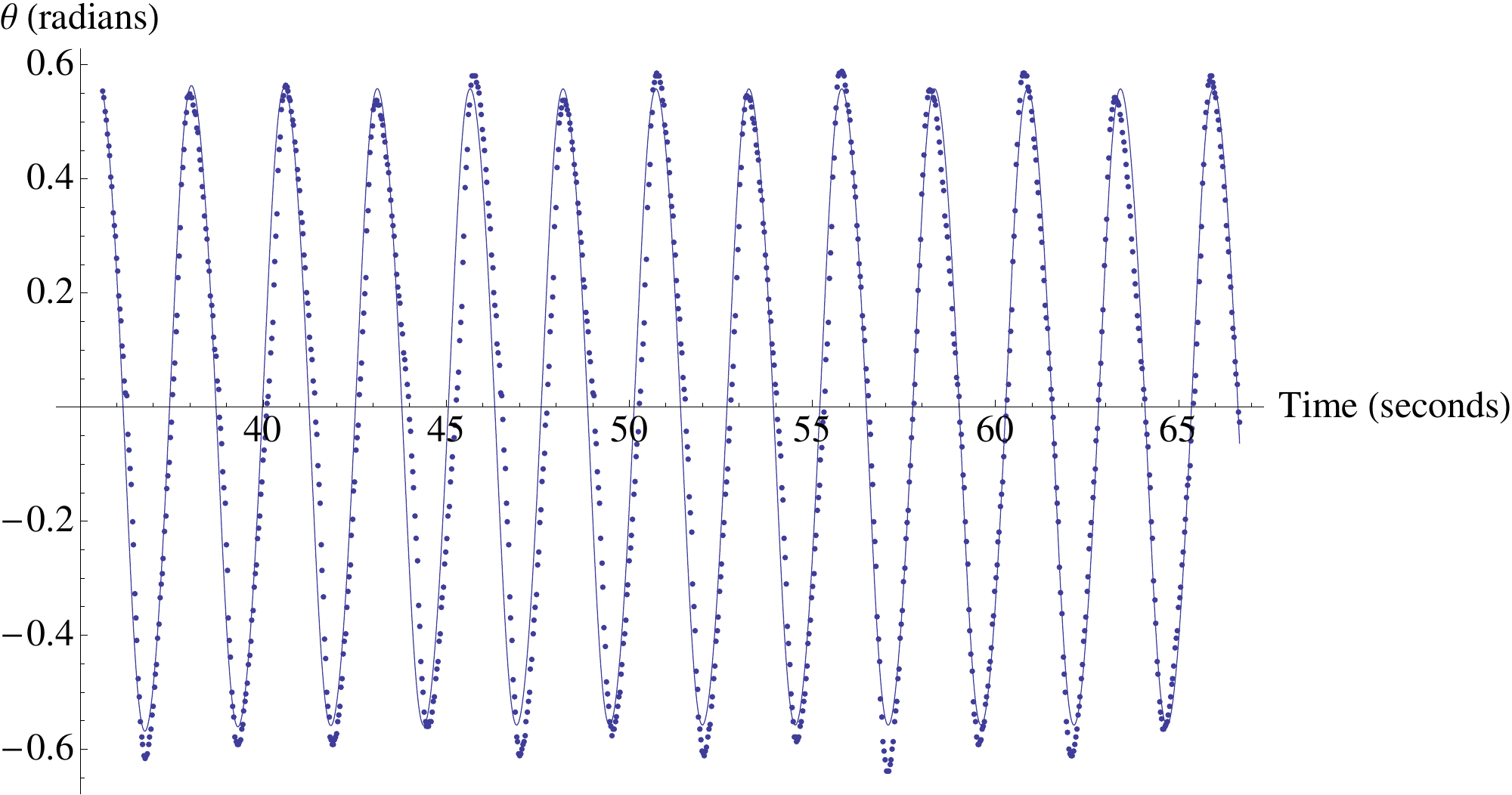}
 \caption{Identification of the model: The continuous graph shows the numerical simulation of the proposed model and the dots show data extracted from the experiments}
 \label{fig:Identification}
\end{figure*}

\begin{figure}[htb]
 \centering
 \includegraphics[width=0.4\textwidth]{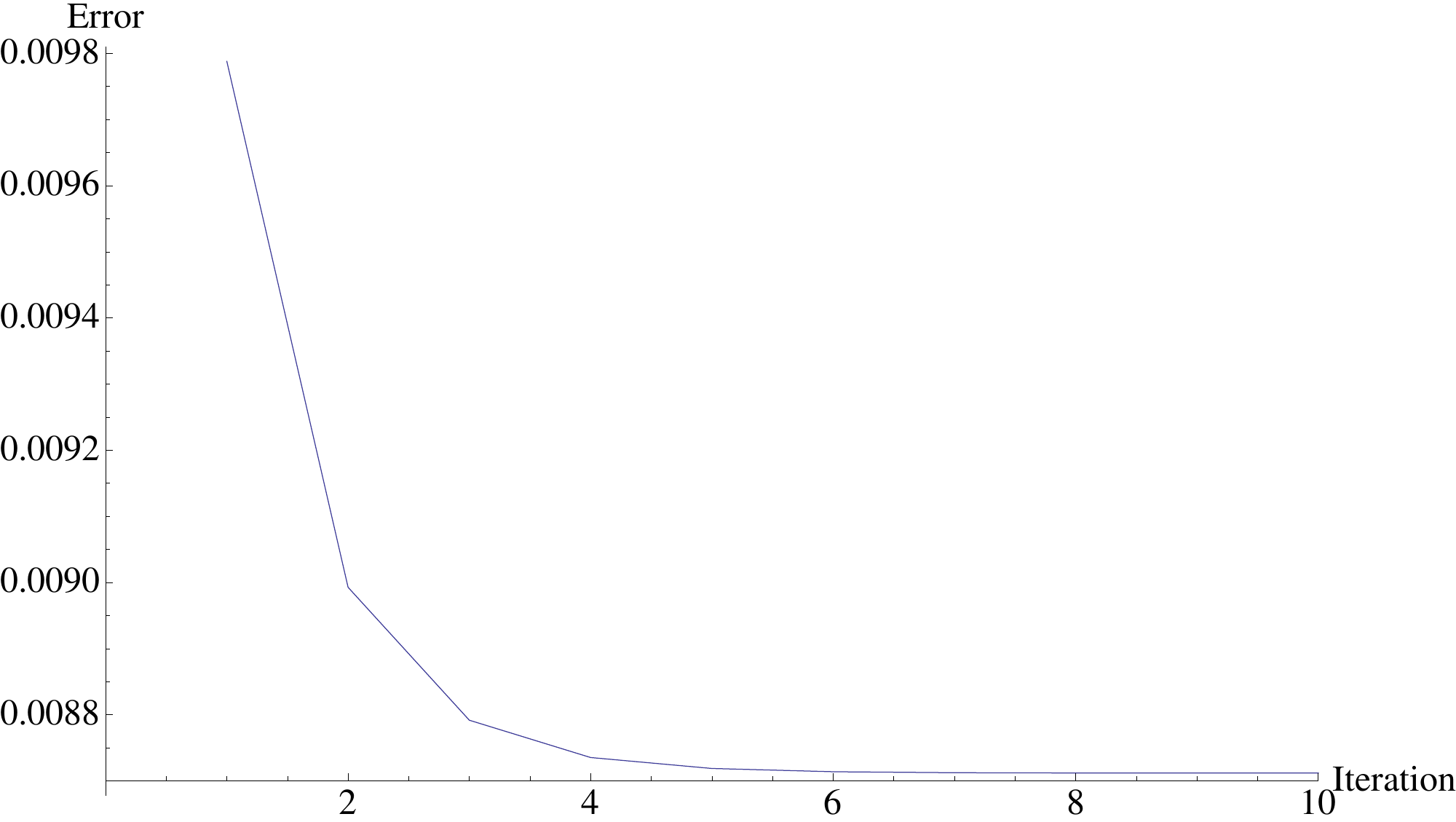}
 \caption{Identification error when one of the parameters is changed during iterations}
 \label{fig:Identification-Error}
\end{figure}

To find out the parameters based on experimental data, we set a metronome to different frequencies and let it oscillate for about 1000 seconds. The data was collected as pairs of time and angle and used for system identification. Using the same initial conditions, a numerical simulation was run with some initial parameters. The error was defined as
\begin{equation}
Error = \frac{\sum\limits_{i = 1}^{N} (\theta(S[i,1]) - S[i,2])}{N}
\end{equation}
in which $\theta(t)$ shows the value of angle in the simulation at time $t$, $S[i,1]$ is the time of capturing the $i$th sample, $S[i,2]$ is the angle value of the $i$th sample and $N$ is the total number of experiment samples. The identification method was to minimize the error between the experimental data and numerical simulation of the differential equations. At each step, two of the parameters were fixed and one parameter was modified in the direction of minimizing the error. When the error was minimized, the parameter was fixed and another one was modified. The process was repeated for several iterations until all the parameters settled in a range that was sufficiently small. For this case we identified the parameters $k$, $\epsilon$ and $\theta_0$. Fig. \ref{fig:Identification-Errors} shows the error value for each of the parameters when the other two parameters are fixed\footnote{Naturally the best value of the parameter on the horizontal axis is the value for which the error on the vertical axis is minimum.}, Fig. \ref{fig:Identification} shows the experiment samples and the proposed model and Fig. \ref{fig:Identification-Error} shows the error when changing the parameters in iterations.

\def \picwidth {0.3}
\begin{figure*}
\centering
\subfigure[$\epsilon$]{
\includegraphics[width=\picwidth\textwidth]
{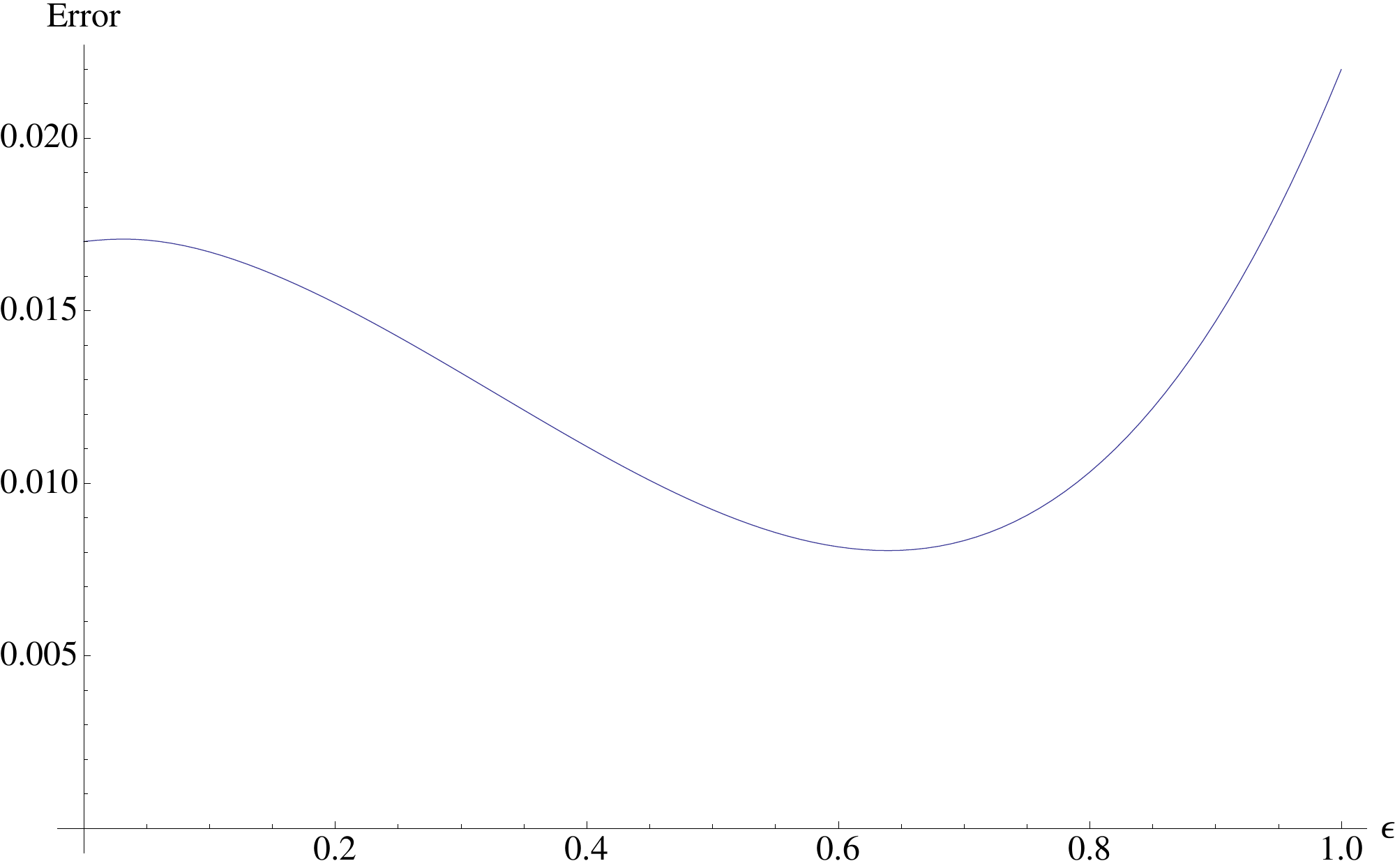}
}
\subfigure[$k$]{
\includegraphics[width=\picwidth\textwidth]
{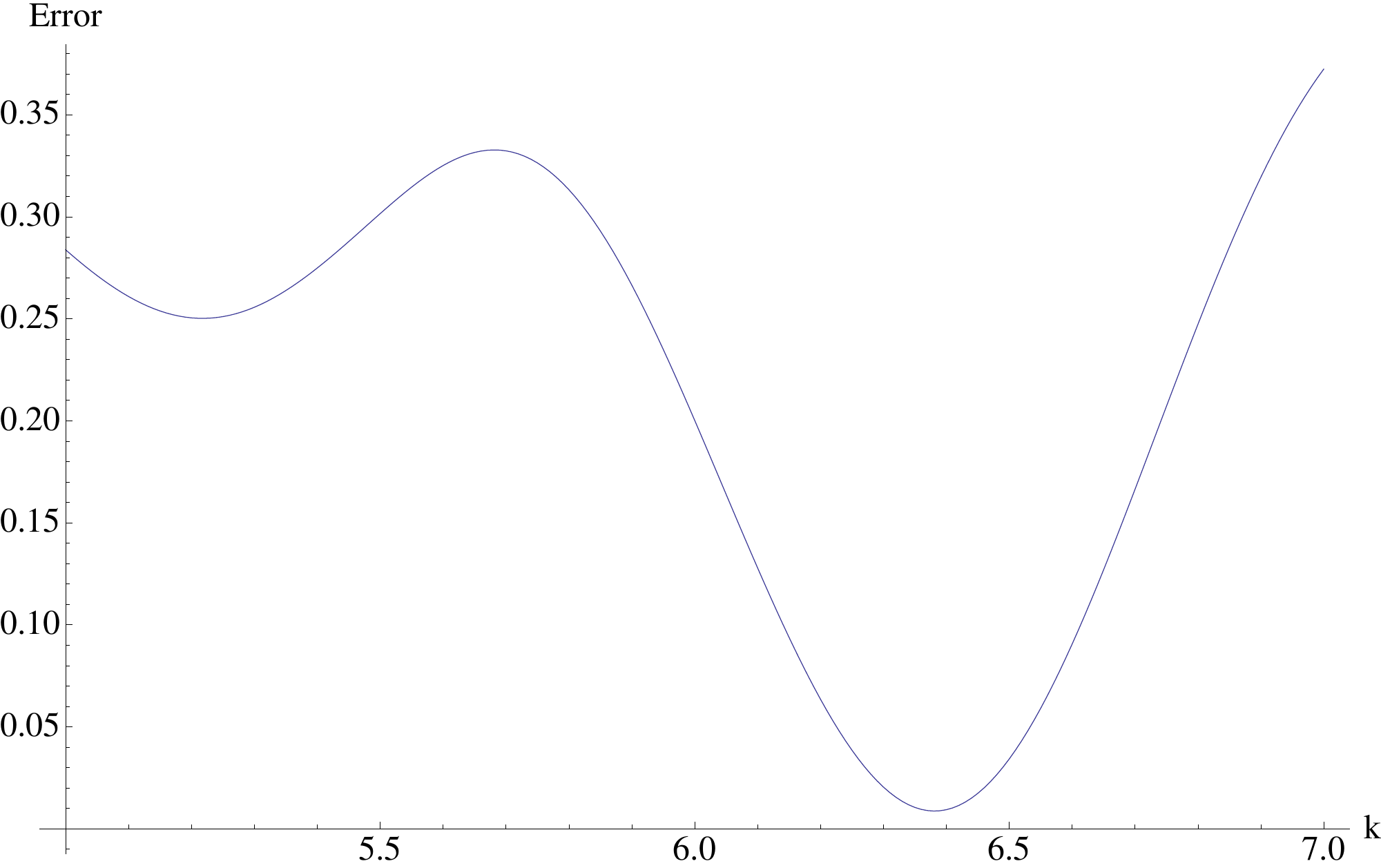}
}
\subfigure[$\theta_0$]{
\includegraphics[width=\picwidth\textwidth]
{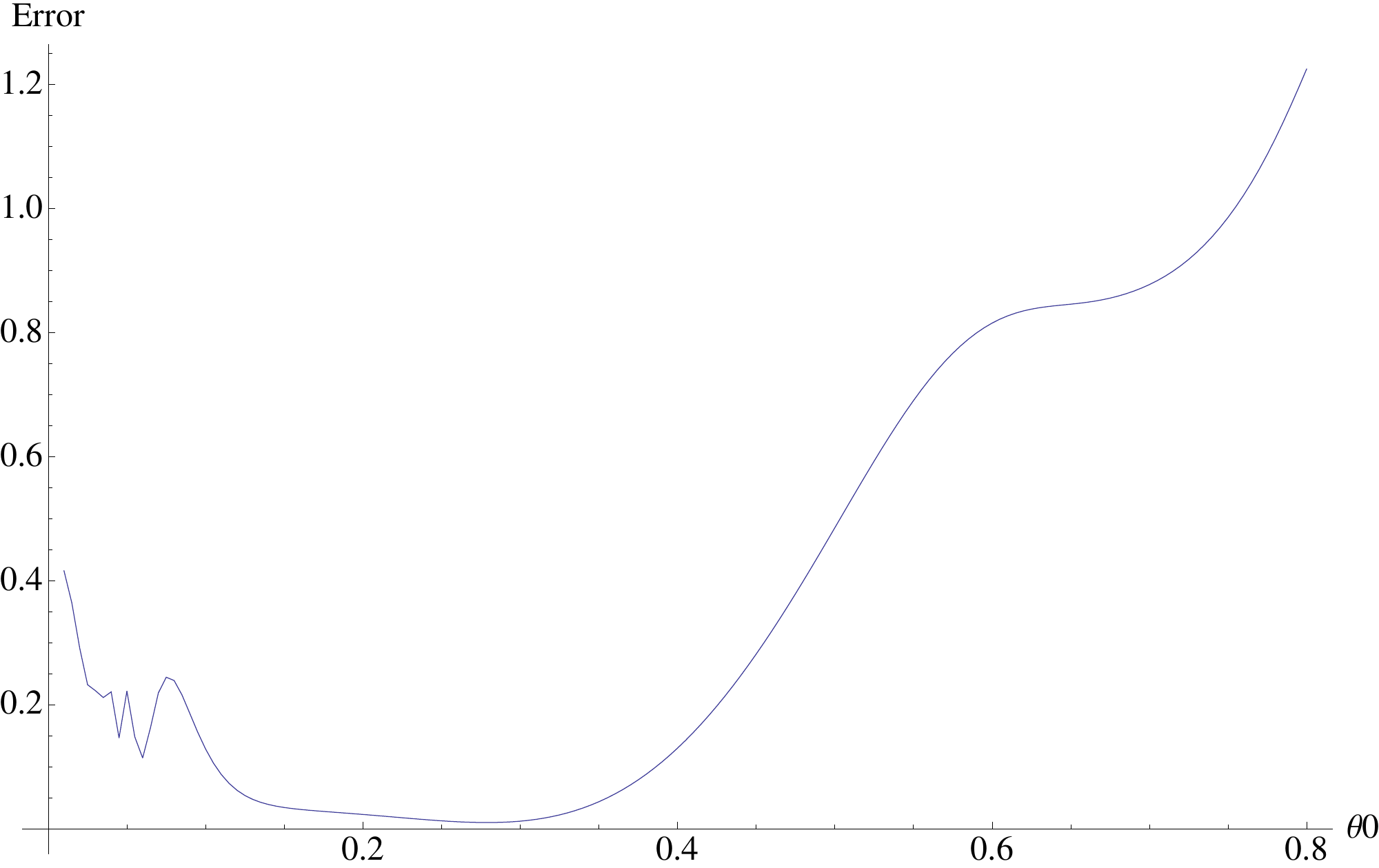}
}
\caption{Identification Errors for three parameters}
\label{fig:Identification-Errors}
\end{figure*}

\clearpage
\clearpage
\section{Synchronization}
\label{sec-synchronization}


Synchronization is a ubiquitous phenomenon everywhere and it plays an important role in many phenomena. We start by reviewing some literature about oscillation, coupled oscillators, synchronization and some related areas.


\subsection{Oscillation and Coupled Oscillators}

The concept of self-oscillations was first proposed by Andronov, Khaikin and Vitt in 1937 \cite{AK-TO-1937} (for the English version see \cite{AVK-TO-1987}). A self-oscillating system is considered as ``an apparatus which produces a periodic process at the expense of a non-periodic source of energy.'' Appleton \cite{A-ASTO-1922} and van der Pol \cite{V-TAFFTV-1920, V-FOCNLR-1927} set experiments with electric circuits while they were studying the reception of radio signals with electric circuits with triodes. In a relaxed self-sustained oscillator, although the motion of a point in the phase plane might be non-uniform, but the growth of the phase in time is still uniform \cite{PRK-SUCNLS-2003}.
%
%

The interaction between the organ pipes was studied by Rayleigh \cite{R-TS-1896}. His experiment involved two organ pipes which had close peach and were placed close to each other. The pipes together sounded in perfect unison. Another investigation is done in \cite{AAB-SSS-2009} in which one pipe was substituted by an electric speaker. The authors observed that even minute driving signals forced the pipe to synchronization.


\subsection{Synchronization}

The word synchronous originates from the Greek words \emph{chronos} and \emph{syn} which means ``sharing the same time'' \cite{PRK-SUCNLS-2003} in which synchronization is considered as a complex dynamical system rather than a state. Synchronization is discussed in \cite{S-SESSO-2003}. It can result from an interaction between systems \cite{FY-STSMCOS-1983} or sub-systems \cite{PC-SCS-1990,PC-DSCS-1991,VLL-SRCS-1992}.

Synchronization of two coupled escapement-driven pendulum clocks was investigated in \cite{S-STCEDPC-2006}. Blekhman \cite{B-SST-1988} discusses observations of the Dutch researcher Christian Huygens \footnote{Christian Huygens is probably the first scientist who observed and described the synchronization phenomenon.} and presents the results of a laboratory reproduction and a theoretical analysis of oscillators coupled through a common supporting frame. At first, Huygens suspected the ``sympathy'' between the clocks was due to induced air currents, but eventually concluded that the cause was the ``imperceptible movements'' of the common supporting structure \cite{BSRW-HC-2002}. In the reproduction of Huygens experiment, the anti-phase state was the only type of synchronization that was observed \cite{BSRW-HC-2002}.

The first observations of synchronization in electronic tube generators were done by Eccles \cite{E-MPCRFMASEMO-1921, E-SWL-1923}. He considered the problem of creating a precision clock and the transmission of naval signals. In \cite{NRBVB-SOPSMHC-2000} the authors experimentally studied rhythmic hand clapping. They consider a conflict between average noise intensity and synchronization. They discuss a mechanism of hand clapping period doubling by individuals that helps the group achieve synchronization. Their results offer a novel route to synchronization, not observed in physics or biological systems by the time of publication.

Synchronization only happens in self-sustained systems \cite{RP-SPCCLCO-2003}. The authors describe self-sustained oscillators mathematically as an autonomous (i.e. without explicit time dependence) nonlinear dynamical systems. Phase of an oscillator is considered neutrally stable. This means that a slight perturbation can change the phase while the amplitude is stable and is not affected by external perturbations. This property provides the oscillator the ability to synchronize \cite{RP-SPCCLCO-2003}.

Synchronization of two metronomes is investigated in \cite{KLNP-STM-2007}. The authors address the problem of analytical study of in-phase synchronization for the model of two metronomes an the common support proposed in \cite{P-SM-2002}. The authors proved the existence of an in-phase regime and they proved that for when the angle difference of metronomes was zero ($\phi_1 - \phi_2 = 0$), the sum of angles ($\phi_1 + \phi_2$) has a periodic regime.

The synchronization of an array of clocks hanging from an elastically fixed horizontal beam is studied in \cite{CPSK-CSNHC-2009}. The beam is considered as a rigid body connected to a spring and a damper. Different types of synchronization were observed: Symmetrical Synchronization, Complete Synchronization and De-synchronous Behavior. Synchronization of two and more metronomes is considered in \cite{P-SM-2002}. Synchronization of coupled mechanical metronomes is also studied in \cite{UMP-SCDCMM-2009}. The authors study synchronization by means of numerical simulations showing the onset of synchronization for two, three and 100 globally coupled metronomes.

Crowd synchrony on London Millennium Bridge \cite{DFFLLRW-LMF-2001} is investigated in \cite{SAMEO-CSMB-2005}. Sufficient conditions for controlled synchronization of non-linear systems is provided in \cite{P-SASSPS-1997}. An attempt to provide a general formalism for synchronization in dynamical systems is shown in \cite{BFNP-OSSCS-1997}. Frequency and coordinate synchronizations are considered in that article. It is mentioned that synchronization as a phenomenon should be considered in context and depends on the view. A system showing synchronization viewed from a view point, might not seem having synchronization from another point of view.

A sample of synchronization is observed in various species. For instance, banded mongoose groups show high degree of birth synchrony to avoid the negative effects of competition with other females \cite{HBC-RCEEBSCM-2010}. Synchronization of time-delayed systems is discussed in \cite{CK-STDS-2007}. A brief introduction to the theory of synchronization of self-sustained oscillators is presented in \cite{PRK-PSRCS-2000}.

Synchronization is investigated in other fields such as Small-World networks \cite{BP-SSWS-2002, CHAHB-SEWCN-2005, TMK-STIPODSWN-2010}, weighted complex networks \cite{LWLF-SWCNHS-2006, CHAB-SWCN-2006, CHMB-DMESCWN-2006} and dynamical networks \cite{BHCAKP-SDNEACG-2006}. Phase synchronization of weakly coupled self-sustained chaotic oscillators is investigated in \cite{RPK-PSCO-1996}. The exact mechanisms of generation of epileptic seizures in human brains is still uncertain. Nevertheless, it is widely accepted that an abnormal synchronization of firing neurons causes epileptic seizures. To investigate this problem, phase synchronization between different regions of the brain is measured \cite{MKADLE-ESPDS-2003, LBHKRSW-SPHEBN-2009}.


\subsection{Model}

A simple model of \emph{coupled oscillators} is given by \cite{S-NDCWAPBCE-2001}:
\begin{equation}
\begin{split}
 \dot{\theta_1} &= \omega_1 + K_1 \sin{(\theta_2 - \theta_1)}
\\
 \dot{\theta_2} &= \omega_2 + K_2 \sin{(\theta_1 - \theta_2)}
\end{split}
\end{equation}
in which $\theta_1$ and $\theta_2$ are the phases of the oscillators and $\omega_1$ and $\omega_2$ are the natural frequencies. The two parameters $K_1$ and $K_2$ determine the amount of dependency between the two oscillators (and show that how much they are affected from the other one). For an uncoupled system, we have $K_1 = K_2 = 0$ and hence the equations are reduced to
\begin{equation}
\begin{split}
 \dot{\theta_1} &= \omega_1
\\
 \dot{\theta_2} &= \omega_2
\end{split}
\end{equation}
When the oscillators are coupled, there is a phase difference $\phi$ between the two oscillators
\begin{equation}
 \phi = \theta_1 - \theta_2
\end{equation}
So, we have
\begin{equation}
\begin{split}
 \dot{\phi} &= \dot{\theta_2} - \dot{\theta_1}
\\
            &= \omega_1 - \omega_2 - (K_1 + K_2) \sin{\phi}
\end{split}
\end{equation}
in which $\omega_1 - \omega_2$ is also called \emph{Frequency detuning} \cite{PRK-SUCNLS-2003}.

Different settings and sets of parameters were used to study synchronization. Some of them and the results are described in this section.


\subsection{Oscillators on a Freely Moving Base}

\begin{figure}
 \centering
 \includegraphics[width=0.45\textwidth]{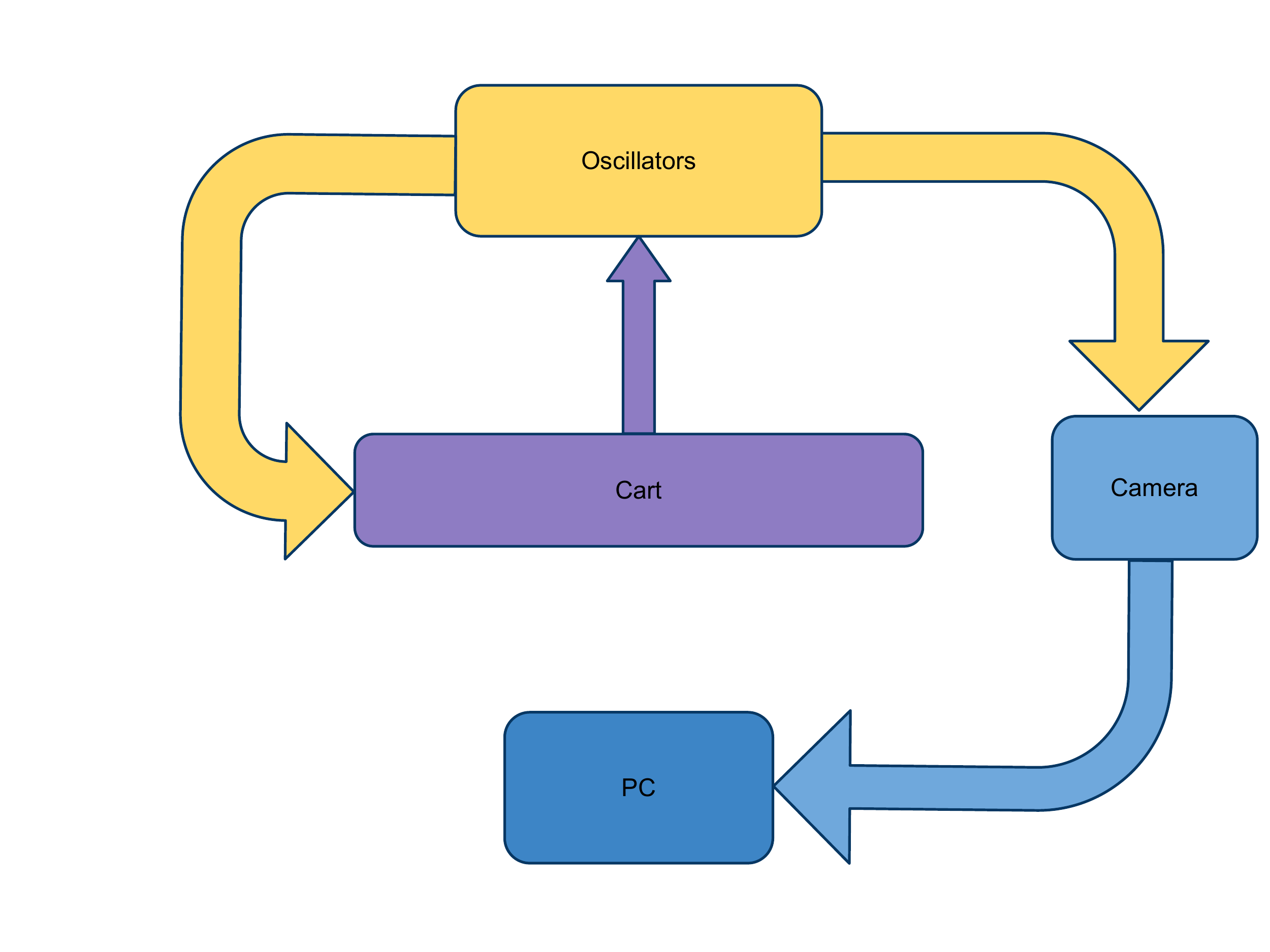}
 \caption{The schematic of the system for two metronomes placed on a freely moving base. The metronomes move the base and the movements of the base affect the metronomes. The camera and the PC are only for observation.}
 \label{fig:20101008-1925-LifePassiveSynchDiagram}
\end{figure}

When the oscillators are placed on a freely moving base(a sample is shown in Fig. \ref{fig:20100920-1650}), they have the possibility of synchronization. A schematic of the settings is shown in Fig. \ref{fig:20101008-1925-LifePassiveSynchDiagram}.
The equations of motion in this case are as follows \cite{P-SM-2002}
\begin{equation}
\begin{split}
\frac{d^2 \theta_1}{d \tau^2} + (1 + \Delta) \sin{(\theta_1)} + \mu \left[\left(\frac{\theta_1}{\theta_0}\right)^2 - 1\right] \frac{d \theta_1}{d \tau}
\\
- \beta \cos{(\theta_1)} \frac{d^2}{d \tau^2} (\sin{(\theta_1)} + \sin{(\theta_2)}) = 0
\\
\frac{d^2 \theta_2}{d \tau^2} + (1 - \Delta) \sin{(\theta_2)} + \mu \left[\left(\frac{\theta_2}{\theta_0}\right)^2 - 1\right] \frac{d \theta_2}{d \tau}
\\
- \beta \cos{(\theta_2)} \frac{d^2}{d \tau^2} (\sin{(\theta_1)} + \sin{(\theta_2)}) = 0
\end{split}
\end{equation}
in which $\tau = \omega t$ is a dimensionless time variable, $\omega^2 = \frac{m r_{c.m.} g}{I}$ is the square of the average angular frequency of the uncoupled, $\Delta \approx \frac{\omega_1 - \omega_2}{\omega}$ is the relative frequency difference between the oscillators, small amplitude oscillator without damping or driving, $\theta_0$ is the van der Pol term and $\beta$ is the coupling parameter
\begin{equation}
\beta = \left( \frac{m r_{c.m.}}{M + 2 m} \right) \left( \frac{r_{c.m.} m}{I} \right)
\end{equation}
in which $M$ is the mass of the base, $I$ is the moment of inertia of the pendulum, $m$ is the mass of the pendulum (masses of the pendulums are considered to be the same) and $r_{c.m.}$ is the distance of the pendulum's center of mass from the pivot point. For low damping of the coupling medium we may observe in-phase synchronization and when the damping of the coupling medium is high, we can observe anti-phase synchronization.


\subsection{Synchronization on a Freely Moving Base}

The metronomes were set to various frequencies and then placed on the freely moving base. The phase differences were observed for each frequency. The experimental results are shown in Fig. \ref{fig:x1-minus-x3}.
\def \picwidth {0.4}
\begin{figure}[htb]
\centering
\subfigure[$f_1 = f_2 = 1.6 Hz$ (192 beats/min)]{
\includegraphics[width=\picwidth\textwidth]
{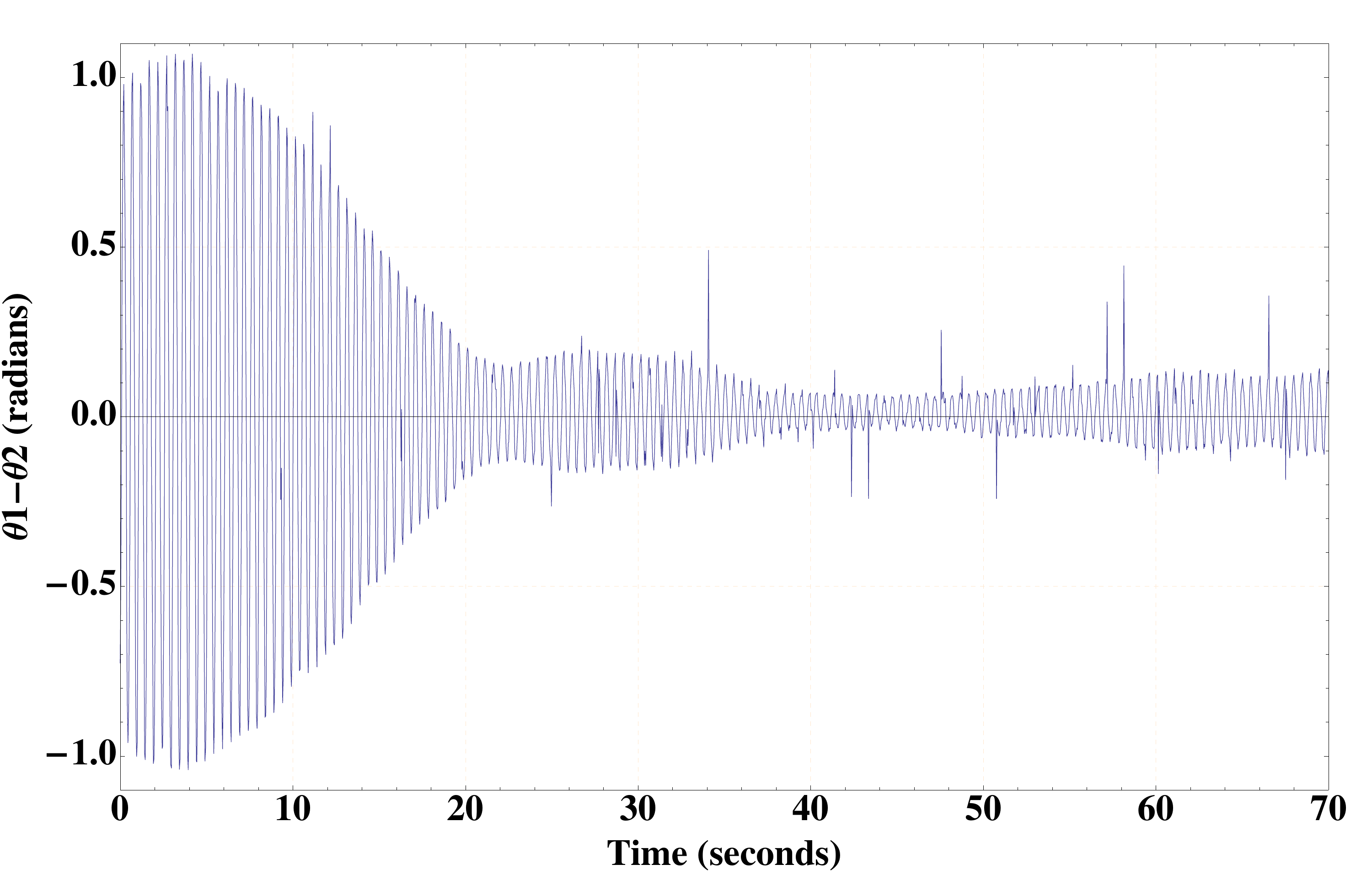}
}
\subfigure[$f_1 = f_2 \approx 1.73 Hz$ (208 beats/min)]{
\includegraphics[width=\picwidth\textwidth]
{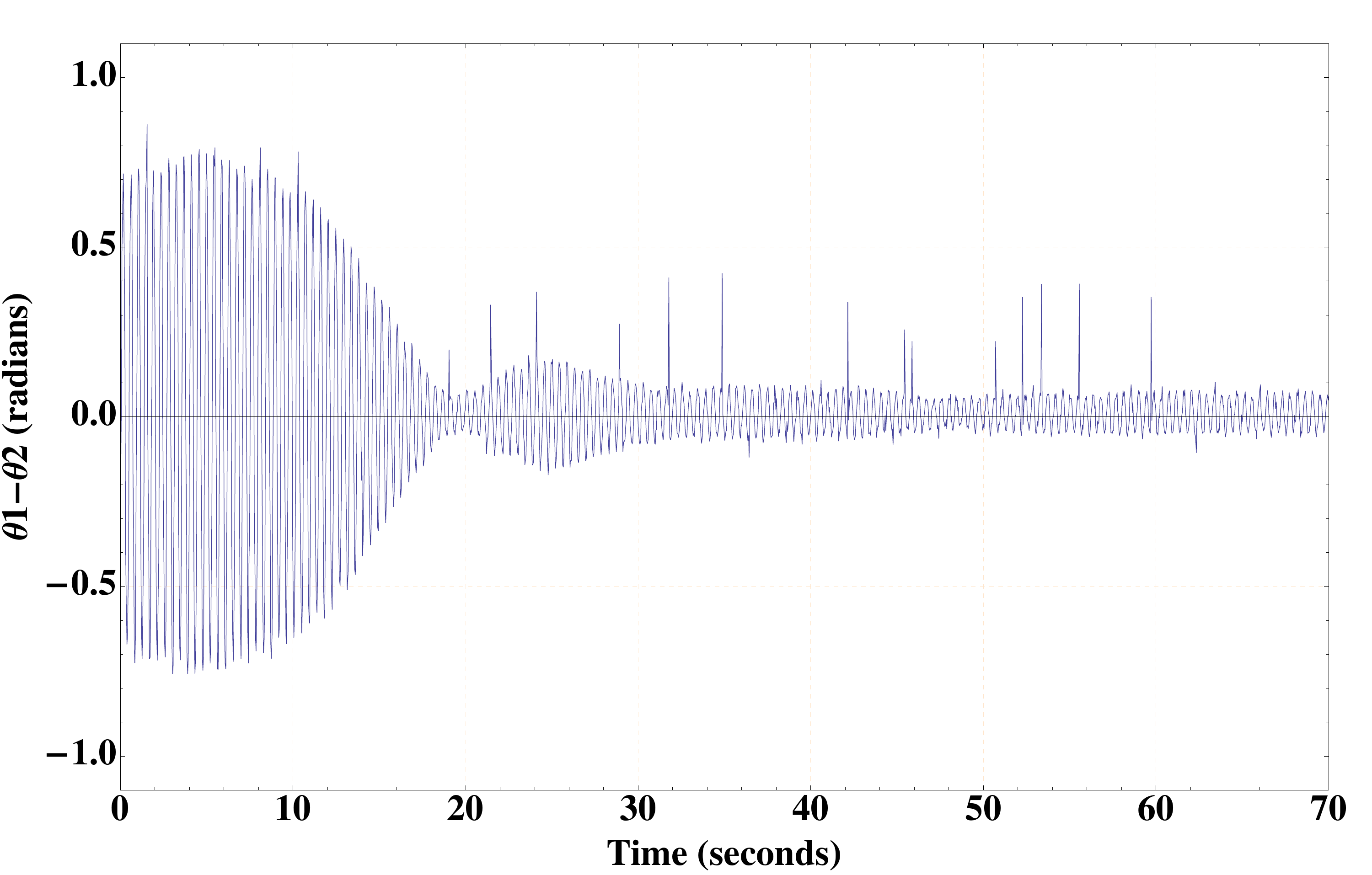}
}
\caption{Phase difference of two metronomes placed on a freely moving base in the experiments}
\label{fig:x1-minus-x3}
\end{figure}


\end{document}